\documentclass[fleqn,twoside]{article}
	
\usepackage{insa}
\usepackage{graphicx}
\newcommand{\mean}[1]{\left\langle #1 \right\rangle}
\newcommand{\be}{\begin{eqnarray*}}
\newcommand{\ee}{\end{eqnarray*}}
\newcommand{\beq}{\begin{equation}}
\newcommand{\eeq}{\end{equation}}
\def\ltap{\ \raisebox{-.4ex}{\rlap{$\sim$}} \raisebox{.4ex}{$<$}\ }
\def\gtap{\ \raisebox{-.4ex}{\rlap{$\sim$}} \raisebox{.4ex}{$>$}\ }

\title{\vskip -2.5cm \hfill {\normalsize TIFR/TH/08-55}\\ \vskip 2.0cm
Heavy Ions at LHC: A Quest for Quark-Gluon Plasma} 

\author{Rajeev S. Bhalerao and Rajiv V.\ Gavai \address{Department of 
Theoretical Physics, Tata Institute of Fundamental Research,\\ Homi Bhabha
Road, Mumbai 400005, India.  }}
       
\begin{document}
\thispagestyle{empty}
\begin{abstract}
Quantum Chromo Dynamics (QCD), the theory of strong interactions, predicts a
transition of the usual matter to a new phase of matter, called Quark-Gluon
Plasma (QGP), at sufficiently high temperatures. The non-perturbative technique
of defining a theory on a space-time lattice has been used to obtain this and
other predictions about the nature of QGP.  Heavy ion collisions at the Large
Hadron Collider in CERN can potentially test these predictions and thereby test
our theoretical understanding of confinement.  This brief review aims at
providing a glimpse of both these aspects of QGP.

\end{abstract}

\maketitle

\section{Introduction}
 
There are two very commonly quoted motivations for the upcoming Large
Hadron Collider (LHC) at CERN in Geneva, the center of attraction for the
articles in this volume. Perhaps the primary one is that LHC will provide
us a key to understand the origin of the visible mass of our Universe.
This alludes to that fact that our standard model(SM) of particle
interactions has to start with matter in the form of massless quarks and
leptons.  The famous Higgs mechanism \cite{hig} of spontaneous breaking of
gauge symmetries provides masses to them, and the carriers of the weak
force, namely $W^\pm$,$Z$.  LHC is widely expected to discover the Higgs
boson which is tied with this mechanism.  The other motivation rests on the
fact that the standard model has been well understood due to the many
impressive precision tests carried out in many experiments, including those
at the Large Electron Positron (LEP) at CERN and the Tevatron at the
Fermilab in the USA.  However, new physics beyond the standard model (BSM)
has to exist \cite{pdg} since SM contains many, at least 19, arbitrary
parameters and thus cannot be the final theory.  Indeed, it is even hoped
that LHC may provide us not only a glimpse of the BSM physics, but it will
hopefully also explain the origin of the mass of the dark matter in
the Universe.

While these motivations are largely correct, there are certain
oversimplifications in them, leading to a few misconceptions, especially in
the popular media.  First of all, even if the expected Higgs particle is
actually discovered, the origin of the mass of up/down$(u/d)$ quarks can be
claimed to be understood only after it is also established that the Higgs
particle couples to them with a strength of $\sim 10^{-6}$, not an easily
achievable goal at LHC.  Indeed, one may as well need an electron-positron
collider to establish this in the post-LHC era.  Moreover, the protons and
neutrons, which make up most of the visible mass in our Universe, have each
a much larger, almost a factor of 100 larger, mass than the sum of the masses of
their constituent $u/d$ quarks.  Therefore, the understanding of the
visible mass of the Universe will emerge from the efforts to figure out why
protons/neutrons have such large binding energies.  Starting from molecules
to atoms and nuclei, we are accustomed to the idea that the interactions
which bind the respective constituents give rise to binding energies much
smaller, less than even a per cent.   This has given rise to the very
successful idea of treating these interactions perturbatively as an
expansion in the strength of the interaction.  As we shall see below, one
needs new suitable techniques to investigate these large binding energies,
in Quantum Chromo Dynamics (QCD), the theory of interactions of quarks with
gluons, the carriers of the strong force.

As may be seen from the articles in this volume itself, QCD is an integral
part of our standard model of particle and their interactions. From various
experiments in the past, it is well known that quarks carry both flavour
quantum numbers such as, electric charge or strangeness, as well as colour:
they transform as a triplet under the colour $SU(3)$ group.  As in the case
of electric charge, the colour charge is also mediated by massless vector
particles, gluons.  Structurally, the theory of quark-gluon interactions,
QCD, looks very similar to that of electron-photon interactions, QED.  A
key difference though is that  there are eight gluons which themselves
carry colour charge, transforming as an octet under $SU(3)$-colour group.
Consequently, gluons can interact amongst themselves.  Furthermore, the QED
coupling is rather small at the scales we probe, being 1/137, whereas the
smallest measured QCD coupling, $\alpha_s$, is about 0.12.  In fact, more
often, one has to deal with $\alpha_s = 0.3$ or so and it is $ \gtap 1$ in
the bound states like proton or neutron.  QCD exhibits a 
much richer structure  and a variety of phenomena as a result of this large
$\alpha_s$. Quark confinement and dynamical chiral symmetry breaking can be
named as typical examples.  A lack of observation of quarks in experiments
led to the hypothesis that quarks are permanently confined in the hadrons,
i.e, protons or pions whereas the lightness of pions compared to protons is
expected to be understood as the phenomenon of dynamical breaking of the
chiral symmetry by the vacuum.  QCD as the theory of strong interactions
has to explain these phenomena.  
Since, QCD is too complex, simple models based on
underlying symmetries are often employed to account for its
non-perturbative aspects.  Indeed, most, if not all, of the ``precision
tests'' are either performed experimentally only at small coupling,
$\alpha_s$, corresponding to rather rare events, or employ the simple
QCD-based models.   The latter are in many cases possible weak links in the
precision tests  of the standard model : physics beyond standard model may
even show up in non-perturbative QCD beyond these models.  We need to look
for it and rule out such a mundane possibility for BSM-physics in order to
be sure that other exotic possibilities are indeed worth looking for.
Thus, non-perturbative techniques are needed for real precision tests of QCD.
As a glaring example, let me mention that the easiest precise measurement
at LHC will perhaps be the total proton-proton cross section at 14 TeV.
The current best theoretical prediction for it is \cite{land} $\sigma^{\rm
tot} = 125 \pm 25 $ mb !   As explained in \cite{land}, one uses the
so-called Regge Models to arrive at it, and one such model can even
explain the currently observed $Q^2$-variation of the structure function of
proton, $F_2$, as well.  Recall that a key cornerstone for establishing QCD
as the theory of strong interaction is this $Q^2$ variation.

While obtaining a reliable prediction for the above cross section from QCD
still seems far away,  a non-perturbative technique does exist today to
obtain other quantities, such as the decay constants or the weak matrix
elements, from QCD using first principles, and these could still provide
non-perturbative precision tests of the standard model.  QCD defined on a
space-time lattice is such a tool.  Not only does it explain many of the
above mentioned phenomena but it provides quantitative estimates of many
physical observables.  Furthermore, the {\em same} techniques of lattice
QCD lead to spectacular predictions for the behaviour of matter under
extreme conditions.  Thus, lattice QCD predicts the existence of a new
phase, called Quark-Gluon Plasma (QGP) at sufficiently high temperature,
and a phase transition of the strongly interacting matter of protons,
neutrons and pions to the new phase QGP at high enough temperature.  The
dynamically broken chiral symmetry of QCD at low temperatures in our world
is expected to be restored in the QGP phase, `melting' away the constituent
mass of the light quarks acquired due to interactions.

Our Universe ought to have existed in such a phase a few microseconds after
the Big Bang, and about 20 microseconds later the phase transition to the
normal hadrons like protons, neutrons and pions ought to have taken place
in it.  Whether there are any imprints of this phase transition on the
astronomical objects observed today depends on the nature of the phase
transition.  There have been speculations of stars with strange matter,
consisting of neutral baryons made from an up, down and a strange quark
each. Similarly attempts have been made to study the influence of such a
phase transition on the Big Bang Nucleosynthesis.  More excitingly, the
Large Hadron Collider (LHC) itself will provide us with an opportunity to
create these Early Universe-like conditions of high energy densities, or
equivalently high temperature, in the laboratory in its proposed heavy ion
collisions of Lead on Lead at 5.5 TeV colliding energy.  Heavy ion
collisions at relativistically high energy have had an illustrious past,
and even more impactful present.  Early such experiments were made at the
SPS collider in CERN, Geneva at a colliding energy of 17 GeV per nucleon in
the center of mass (cm) frame.  The relativistic heavy ion collider (RHIC)
has been operative in BNL, New York, since a last few years and has produced heavy ion
collision data for a variety of ions, Deuterium (D), Copper(Cu), and Gold
(Au), at a spectrum of energies, 62--200 GeV per nucleon in the cm frame.
Experiments at LHC will thus see a further jump in the colliding energy by a
factor of about 30.   It is hoped that this will offer us cleanest
environment yet for investigating the physics of quark-gluon plasma.

In this short review, we shall attempt to provide a glimpse of how lattice
QCD leads to QGP and predicts many of its properties as well as those of
the corresponding phase transition and how the heavy ion collision experiments 
amazingly provide us an opportunity to produce QGP in a laboratory,
including the expectations of what we may observe at LHC.

\section{QGP from Lattice QCD}

In order to understand and appreciate the fundamental importance of
attempts to discover QGP at the LHC, let us first review the basics of
lattice QCD and why it facilitates a truly reliable treatment
of non-perturbative
physics.  In the process, we shall also see why essentially the {\em same}
tested technique for obtaining, say, the hadron masses, comes into play for
predicting new phases or phase transitions.

\subsection{Basic Lattice QCD}

\begin{figure}[htb]
\includegraphics[scale=0.50]{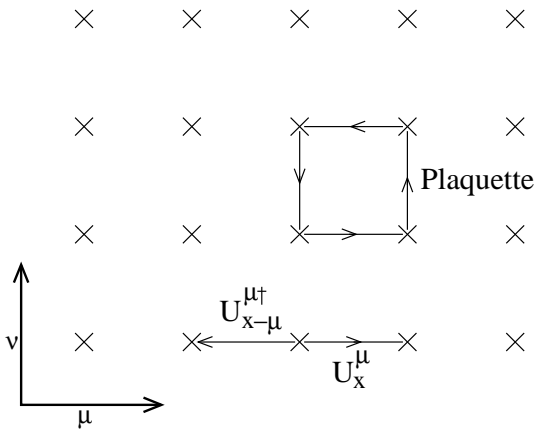}
\vskip -0.5 cm
\caption{Quark and gluon fields on a space-time lattice.}
\vskip -0.5 cm
\label{fig:one}
\end{figure}

Lattice field theory is defined by discretizing the space-time. The (inverse
of the) lattice spacing $a$ acts as the ultra-violet cut-off needed to tame
the divergences in a quantum field theory. One places the anti-commuting
quark fields $\psi(x)$, and $\bar \psi(x)$ on lattice sites whereas the
gluon fields reside on the links, as shown in Figure \ref{fig:one}.
A directed link from site $x$ in the positive direction $\hat \mu$ is
associated with the gluon field $U^\mu_x$, while the link to the site
$x - \hat \mu$ in the opposite direction is $U^\mu_{x - \hat \mu}$. 
A gauge transformation  $V_x \in SU(3)$ rotates the quark field in the colour
space : $\psi'(x) = V_x \psi(x)$. Demanding that the gluon field at the link
$x$ in the direction $\hat \mu$, $U_\mu(x)$, change to $U'_\mu(x) = V_x
U_\mu(x) V_{x+\hat \mu}^{-1}$, ensures that the (discrete) kinetic energy term
of quarks remains invariant under such a gauge transformation. Constructing
gauge actions from closed Wilson loops of the links, like e.g., the smallest
square loop, called plaquette and displayed in Figure \ref{fig:one}, ensures
their gauge invariance.

It turns out that a straightforward discretization of the derivative, given
by [$a \cdot \partial_\mu \psi (x) = \psi(x + a \hat \mu) - \psi(x - a \hat
\mu)$], can be made gauge invariant as shown in the  Figure \ref{fig:one},
where the links end on respective quark fields $\psi$ at the sites.  Thus a
sum over all independent terms of both types shown in Figure \ref{fig:one}
yields the QCD action on the lattice.  However, it leads to the so-called
Fermion Doubling problem : each lattice fermion corresponds to $2^d = 16$
flavours in the continuum limit of $a \to 0$.  Various lattice Fermion
actions, referred to as the Staggered, Wilson, Domain Wall or Overlap
Fermions, have been proposed to alleviate this problem.  In view of their
simplicity and an exact chiral symmetry even on the lattice, the staggered
Fermions have dominated the field of interest for this article, namely
lattice QCD at finite temperature and density.   Briefly, these are single
component Grassmann variables on each site, with the $\gamma$-matrices
replaced by suitably defined sign factors.  They have a $ U(1) \times U(1)$
chiral symmetry and 4 flavours in continuum limit.  An oft-discussed
problem of the staggered Fermions, though, is that two or three light
flavours are not simple to define, and the currently used methods may miss
out on important physics aspects related to anomalies.  It is often argued
that for the bulk thermodynamic properties these issues are likely to be
unimportant.

Typically, for any lattice computation one needs to evaluate the
expectation value of an observable $\Theta$,
\begin{equation}
\langle \Theta(m_v) \rangle = {{ \int \scriptstyle{D} U \exp (-S_G) 
\Theta(m_v)~ {\rm Det}~M(m_s)}
\over { \int \scriptstyle{D} U \exp (-S_G)~ {\rm Det}~M(m_s)} }~~,
\label{therm}
\end{equation}
\noindent 
where $M$ is the Dirac matrix in $x$, colour, spin, flavour space for sea
quarks of mass $m_s$, $S_G$ is the gluonic action, and the observable
$\Theta$ may contain fermion propagators of mass $m_v$.  $S_G \sim 6 \sum
tr U_{plaq}/ g_0^2$, with $g_0$ the bare coupling and $U_{plaq}$ the
product links along a plaquette as shown in Figure \ref{fig:one}.  Amongst
the many methods of evaluation of eq.(\ref{therm}), numerical simulations
stand out due to the ability to achieve the goal of removing the lattice
scaffolding, i.e., taking the continuum limit $a \to 0$.  Using the
two-loop $\beta$-function, it is easy to show that 
\begin{equation}
M \cdot a = \frac{M}{\Lambda}(g^2_0b_0)^{-b_1/2b^2_0} e^{-\frac{1}{2b_0g_0^2}} 
(1+0(g^2_0)~,~
\label{cont}
\end{equation}
defines the way a mass scales $M \cdot a$ on the lattice changes as the bare
coupling $g_0 ( = \sqrt{6/\beta})$ is changed.  Here $b_0$ and $b_1$ are the
universal coefficients of the $\beta$-function.  Typically, one needs larger
and larger lattice sizes as $a \to 0$ in order to keep physical volume fixed.
 
Numerically, the $\langle \Theta \rangle $ is computed by averaging over a
set of configurations $\{ U_\mu(x)\}$ which occur with probability $
\propto \exp(-S_G)\cdot {\rm Det}~M$. Thus the main problem is to generate
the ensembles of such configurations with the desired probability
distribution.  Complexity of evaluation of Det $M$ has lead to various
levels of approximations in the process of generation of configurations:
the quenched approximation consists of sea quark mass, $m_s = \infty$ limit
whereas the full theory should have low sea quark masses: $ m_u = m_d$ with
a moderately heavy strange quark. The computer time required to obtain
results at the same precision increases as the sea quark mass is lowered.

\subsection{Some Results from Lattice QCD}

\begin{figure}[htb]
\hskip 0.2 cm \includegraphics[scale=0.43]{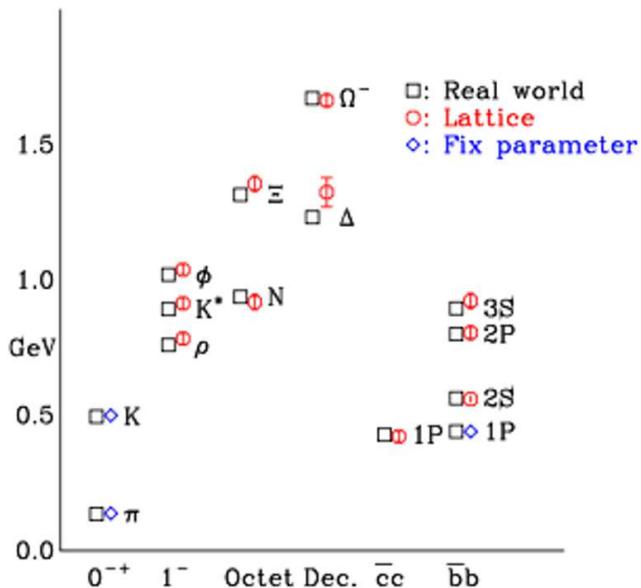}
\vskip -0.5 cm
\caption{Comparison of experimental hadron spectra with lattice results \cite
{milc}.}
\vskip -0.5 cm
\label{fig:two}
\end{figure}

\begin{figure}[htb]
\includegraphics[scale=0.57]{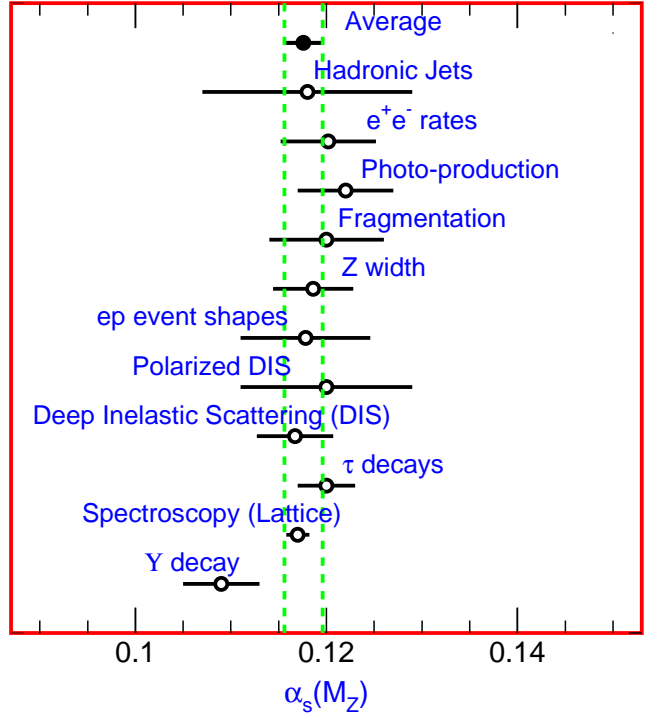}
\vskip -0.5 cm
\caption{Various determinations of $\alpha_s$. From \cite{hpqcd}.}
\vskip -0.5 cm
\label{fig:three}
\end{figure}

A variety of qualitative and quantitative results have been obtained using
the lattice techniques.   It will be both impractical and unnecessary to
review all of them here.  However, in order to appreciate the power of
these techniques, we limit ourselves to providing a glimpse of them for the
staggered fermions; similar, sometimes better in quality/precision, results
have been obtained with the Wilson fermions as well. Figure \ref{fig:two}
shows \cite{milc} the results of the MILC and HPQCD collaborations for the
light as well as heavy hadrons obtained with light sea quarks.  Using the
pion and kaon masses to fix the scales of the corresponding quark masses,
most other particle masses are found to be in good agreement with the
experiment.  Furthermore, the spontaneous breaking of the chiral symmetry
has been demonstrated by many groups since the early days of the lattice
QCD, showing a non-vanishing chiral condensate, $ \langle \bar \psi \psi
\rangle \ne 0$. Moreover, the goldstone nature of the pion has also been
verified by checking that $m_\pi^2 \propto m_u$.  Figure \ref{fig:three}
displays a comparison \cite{hpqcd}  of the lattice determination of the
strong coupling, $\alpha_s(M_Z)$, with other perturbative determinations
from experimental data.  While these results verify that QCD is indeed the
correct theory of the strong interactions, and the lattice technique is the
most reliable quantitative tool to extract its non-perturbative properties,
making new predictions for the experiments is where the real challenges and
excitement lies.  It is very heartening to note that the decay constants of
pseudo-scalar mesons containing a heavy quark were first obtained using
lattice techniques: $f_{D^+} = 201 \pm 3 \pm 17$ MeV and $f_{D_s} = 249 \pm
3 \pm 16$ MeV \cite{milc2}. These have since been measured experimentally
to be $f_{D^+} = 223 \pm 16 \pm 7$ MeV \cite{cleo} and $f_{D_s} = 283 \pm
17 \pm 14$ MeV \cite{babar}, in excellent agreement with the lattice QCD
predictions.

\subsection{Lattice QCD at Nonzero Temperature and Density}

Investigations of QCD under extreme conditions, such as high temperatures
and/or densities, provide a solid platform for its most spectacular
non-perturbative tests.  Since the results from hadron spectroscopy fix the
quark masses as well as the scale $\Lambda_{QCD}$, these tests are even
completely free of any arbitrary parameters.  Based on simple models, which
build in the crucial properties of confinement or chiral symmetry breaking
and allow asymptotically for the free quark gluon gas, one expects phase
transitions to new phases such the Quark-Gluon Plasma or the colour
superconductors.  As we shall see in the next section, the experimental
possibilities of creating the required temperature, and thus the new QGP
phase, exist in the heavy ion collisions at high energies in BNL, New York
and CERN, Geneva. Considering the scale of the entire experimental
enterprise, both in man-years invested and money spent, it seems absolutely
necessary to have a better theoretical foundation for these results
compared to merely relying on simple models.  Fortunately, one can use the
canonical Euclidean field theory formalism for equilibrium thermodynamics
to look for the new phases, and the phase transitions in {\em ab~initio}
calculations from the underlying field theory, i.e., QCD. Indeed,
properties of the QGP phase can be predicted theoretically using the
lattice QCD approach, and tested in the experiments at BNL and CERN.  As a
first principles based and parameter-free approach, Lattice QCD is an ideal
reliable tool to establish the QCD phase diagram and the properties of its
many phases.  While most other basic features of the lattice formalism
required for such an exercise remain the same as in section {\bf 2.1}, a
key difference for simulations at finite temperature is the need of an $
N_s^3 \times N_t$ lattice with the spatial lattice size, $N_s \gg N_t$, the
temporal lattice size for the thermodynamic limit of $V = N_s^3 a^3 \to
\infty$.  The temperature $T = 1/(N_t \cdot a)$ provides the scale to
define the continuum limit : Fixing the transition temperature in physical
(MeV) units and using eq. (\ref{cont}), the continuum limit is obtained by
sending $N_t \to \infty$.
 
The lattice QCD approach has provided information on the transition
temperature, the order of the phase transition, and the equation of state
of QCD matter.  One exploits the symmetries of the theory to construct
order parameters which are then studied as a function of temperature to
look for phase transitions, if any. QCD has two different symmetries in
opposite limits of the quark mass $m_q$.  For $N_f$ flavours of massless
quarks, QCD has $SU(N_f) \times SU(N_f)$ chiral symmetry while for $m_q \to
\infty$, it has a global $Z(3)$ symmetry.  Such symmetries usually imply
zero expectation values for observables which transform nontrivially under
it unless the symmetry is broken spontaneously due to dynamical reasons and
the vacuum transforms nontrivially under it.  Lattice techniques enabled us
to establish that the chiral symmetry is broken spontaneously at low
temperatures, as indicated by its non-vanishing order parameter, the chiral
condensate $\langle \bar \psi \psi \rangle \ne 0$.   Its abrupt restoration
to zero at high temperature will be a signal of a chiral symmetry restoring
phase transition.  Since the chiral condensate can be regarded as an
effective mass of a quark, arising due to QCD interactions, the chiral
transition can be interpreted as thermal effects `melting' this mass.
Similarly, the global $Z(3)$ symmetry breaking can be shown to be
equivalent to a single quark having a finite free energy, i.e., the
existence of a free quark.  A nonzero expectation value for its order
parameter, the Polyakov loop $ \langle L \rangle $, is the a signal for
deconfinement.   Of course, in our world with two light and one moderately
heavy flavours, neither symmetry is exact but these order parameters may
still act as beacons for transitions, depending on how mildly or strongly
broken they are.

\subsection{Results from Lattice QCD at $T \ne 0$.}

The transition temperature $T_c$ can be determined by locating the point of
discontinuity or sudden change in the order parameter as a function of the
temperature (or other external parameter such as density).   Since
numerical results are necessarily obtained on finite lattices, there is an
inevitable rounding which makes the determination of $T_c$ a little tricky.
A lot of work has been done on this question in the statistical mechanics
area and standard finite size scaling techniques exist to pin down $T_c$ as
well as the order of the transition.   Since the early days, numerical
simulations of lattice QCD have progressively tried to approach the real
world of light quarks with vanishing effects from the lattice cut-off.  The
efforts began from the quenched approximation, i.e., QCD without dynamical
quarks, where the deconfinement order parameter $\langle L \rangle$ on
small $N_t$-lattices was used to establish a first order deconfinement
phase transition.   Later QCD with three or more light dynamical quarks was
also shown to have a first order chiral transition.  Recent work on
simulations for QCD with a realistic quark spectrum seems \cite{fodor1} to
rule out a first order chiral transition or a second order transition with
the expected $O(4)$-exponents, but suggests a rapid cross over.
Determination of $T_c$, now the point of sharpest change, is even more
tricky as a result.  The current range for it can be summarized to be 
170-190 MeV.  A value on the lower end of the range was obtained
\cite{fodor2} by using larger $N_t$-lattices while a value at the upper end
was obtained \cite{cheng} using improved action but smaller $N_t$.   There
are other technical differences, such as the physical observable used to
set the scale of lattice QCD, as well. Since the energy density is
proportional to $T^4$, the current uncertainty in the value of $T_c$
translates to a $\sim60$ \% difference in the corresponding energy density 
estimates at $T_c$.  In view of the tremendous impact it has on the
requirements of heavy ion collision experiments, it is hoped that a
narrowing of the range takes place as a result of future lattice QCD work.

\begin{figure}[htb]
\includegraphics[scale=0.70]{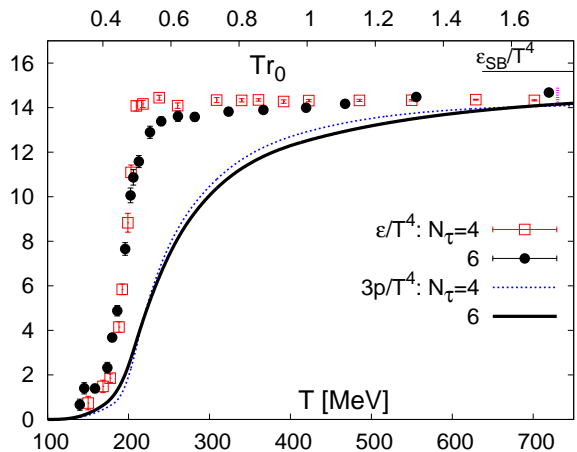}
\vskip -0.5 cm
\caption{Energy density and Pressure from lattice QCD. Taken from \cite{rbrc}.}
\vskip -0.5 cm
\label{fig:four}
\end{figure}

\begin{figure}[htb]
\includegraphics[scale=0.70]{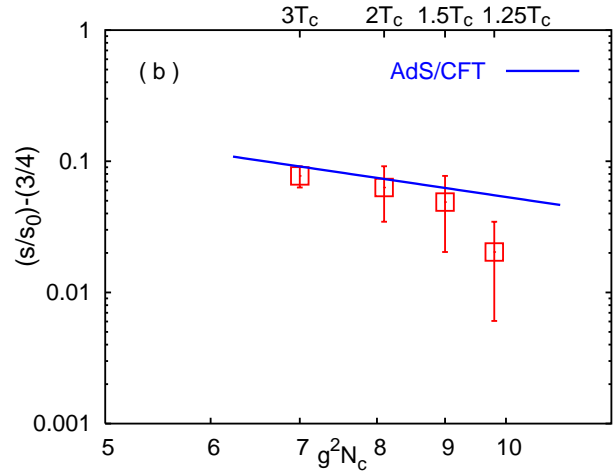}
\vskip -0.5 cm
\caption{Entropy density $s$ (in units of ideal gas entropy $s_0$) as a
function of 't Hooft coupling. From \cite{mgg}.}
\vskip -0.5 cm
\label{fig:five}
\end{figure}

\begin{figure}[htb]
\includegraphics[scale=0.68]{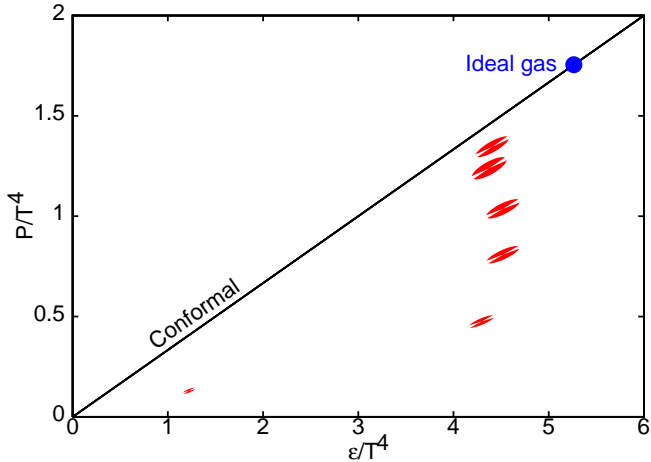}
\vskip -0.5 cm
\caption{Equation of State for (quenched) lattice QCD. Taken from \cite{mgg}.}
\vskip -0.5 cm
\label{fig:six}
\end{figure}

Quantities of thermodynamic interest such as the energy density, or the
pressure or various quark number susceptibilities can be obtained by using the
canonical relations from statistical mechanics.  Thus, 
\begin{equation}
\epsilon = \frac{T^2}{V} \left. \frac {\partial \ln {\cal Z}}{\partial T} 
\right|_{V,\mu}~~{\rm or}~~\chi_B =
\frac{T}{V} \left. \frac {\partial^2 \ln {\cal Z}}{\partial \mu_B^2 } \right|_{V,T}~~,~{\rm etc.}
\label{EnChi}
\end{equation}
\noindent Early results in the quenched QCD showed the existence of 
a QGP phase which has energy density of about 85 \% of the corresponding
ideal gas. The progress since then has been in employing large $N_t$ and
inclusion of light quark loops.  Figure \ref{fig:four} displays recent results
from such efforts.  Obtained on two different lattice sizes, $N_t =4$ and 6
with nearly realistic $u,d$ and $s$ masses, these results also exhibit similar
kind of, $\sim 15 \%$,  deviations from the ideal gas and do seem to hint
towards the lattice cut-off effects to be small .  The spatial volumes are perhaps
not large enough to ensure that the thermodynamic limit is reached.  However,
this question is likely be addressed in near future soon.  The results also
suggest at most a continuous transition or even a rapid cross over; a strong
first order phase transition assumed/constructed in many phenomenological
models seems clearly ruled out.  This has implications for the hydrodynamical
models used to analyse the experimental data: possible mixed state of
quark-gluon plasma and hadronic gas must be short lived, if at all it exists.

From a theoretical perspective investigation of equation of state offers hints
of developing analytic or semianalytic approaches.  Thus conformal invariant
theories are known to yield a variety of predictions for the thermodynamic
quantities using the famous AdS-CFT correspondence. Figure \ref{fig:five} shows
an attempt to confront the entropy density \cite{mgg} for the quenched QCD in
terms of the entropy of the ideal gas with the prediction of  $N=4$ SYM
\cite{GubKle}.  The agreement is impressive, considering the differences of
the underlying theories.  On the other hand, it is really in the stronger
coupling region that it is not as good.  Moreover, resummed weak coupling
perturbation theory approaches seem to perform equally well at the lower
couplings.   Figure \ref{fig:six} shows the results \cite{mgg} for the equation
of state to highlight how conformal QCD really is. The ellipses denote 66\%
error bounds on the measured EOS.  The wedges piercing the ellipses have
average slope $c_s^2$, the speed of sound and the opening half-angle of these
wedges indicate the error in $c_s^2$.   Conformal invariance is indeed violated
significantly in the region close to the transition, with least violation at
the same temperatures where in AdS-CFT prediction does well in Figure 
\ref{fig:five}.

\begin{figure}[htb]
\includegraphics[scale=0.32]{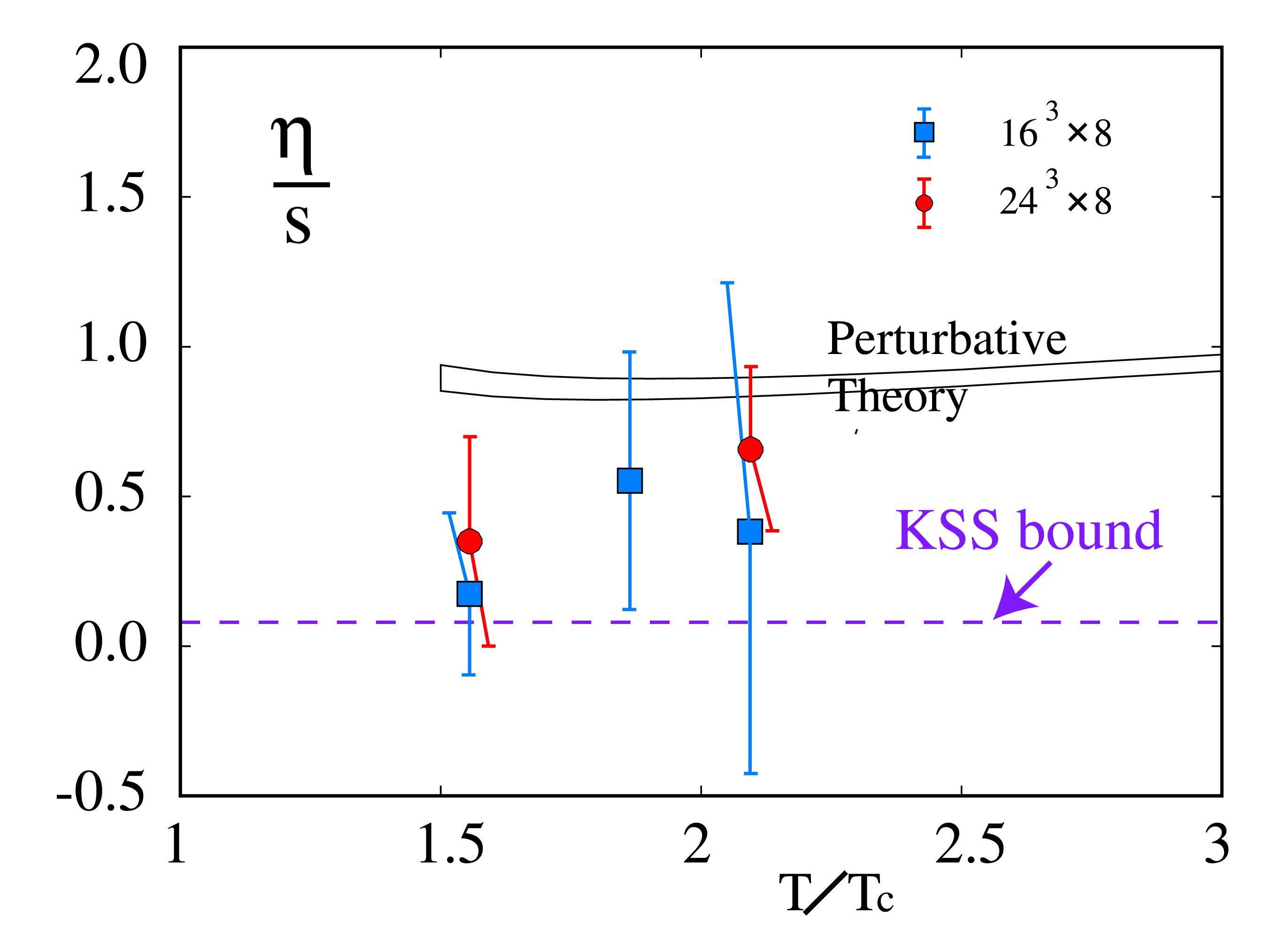}
\vskip -0.5 cm
\caption{Ratio of shear viscosity to entropy in (quenched) QCD vs. 
temperature. Taken from \cite{NaSa}.}
\vskip -0.5 cm
\label{fig:seven}
\end{figure}

Viscosities of the quark-gluon plasma, both the shear ($\eta$) and bulk
($\zeta$), can also be determined using the lattice approach although unlike
the equation of state these determinations need extra ans\"atze some of which
are not universally accepted.  Kubo's linear response theory lays down the
framework to obtain such transport coefficients from certain equilibrium
correlation functions.  In particular, one obtains correlation functions of
energy-momentum tensor using the lattice approach above.  These are, of course,
defined at discrete Matsubara frequencies.  Recall that the simulations at $T
\ne 0$ need lattices with i) periodic boundary conditions and ii) small $N_t$
compared to $N_s$.  The correlation function is thus defined at few discrete
points only.  One then continues it analytically to get the so-called retarded
propagators in real time from which the the $\eta$ and $\zeta$ are obtained in
the zero frequency limit.  Figure \ref{fig:seven} shows the results \cite{NaSa}
in the quenched approximation.  Close to $T_c$, rather small values are
obtained for the ratio of $\eta$ to the entropy density $s$.  These are seen to
be consistent with the famous bound \cite{kss} from AdS-CFT.  As shown in the
Figure, perturbation theory suggests rather large values for this ratio.  These
results have since been refined \cite{meyer} and made more precise but the
general picture remains the same, as do the various theoretical uncertainties
which plague these determinations.  Larger lattices and inclusion of dynamical
quarks will surely reduce some of these in near future.  What is needed though
for a more convincing demonstration of the fact the shear viscosity is indeed
as small as hinted by the experimental data (see the next section) is a better
control over the systematic errors in the analytic continuation. 

Analogous to the baryon number susceptibility, defined in eq. (\ref{EnChi}),
various quark number susceptibilities can be defined by taking derivatives with
the appropriate chemical potential.  These determine the fluctuations in the
given conserved quantum number, say, strangeness.  It has been argued
\cite{GavGup} that under certain assumptions, testable experimentally, the
strange susceptibility can be related to the Wr\'oblewski parameter $\lambda_s$
extracted from the data of heavy ion collisions. Interestingly, lattice QCD
computation in both quenched approximation and full QCD yield a $\lambda_s
(T_c) \simeq 0.4-0.5 $, whereas various experimental results \cite{clm} lead to
a value $0.47 \pm 0.04$.   Taking derivatives with two different chemical
potentials in eq. (\ref{EnChi}), one obtains off-diagonal susceptibilities.
These have the information on flavour correlations.  Such a baryon-strangeness
\cite{kmr} or electric charge-strangeness \cite{GavGup} correlation has been
proposed as a signature for identifying the nature of the high temperature
phase as that of the quark-gluon phase.

\begin{figure}[htb]
\includegraphics[scale=0.66]{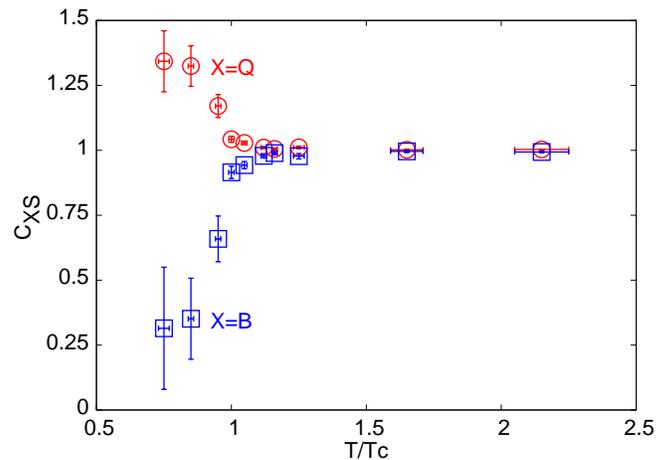}
\vskip -0.5 cm
\caption{ Baryon-Strangeness and Electric charge-Strangeness correlation vs.
temperature \cite{GavGup}.}  
\vskip -0.5 cm
\label{fig:eight}
\end{figure}

\noindent Figure \ref{fig:eight} shows the lattice results for QCD with 2 light
dynamical quarks for both these correlations.  They have been so normalized
that a value unity, as seen in most of the high temperature phase in Figure
\ref{fig:eight}, characterises the existence of quark degrees of freedom with
the appropriately fractional baryon number or charge.  It has been shown that
the correlation  in the low temperature phase are consistent with the hadronic
degrees of freedom.  Indeed, any lack of the expected transition should lead to
much milder temperature dependence as well as a value different from unity for
these correlation functions.  Being ratios of the quark number susceptibilities,
these correlations are robust, both theoretically and experimentally.
Systematic errors  due to lattice cut-off or dynamical quark masses are
therefore very small as are the systematic errors from experimental sources.

\begin{figure}[htb]
\includegraphics[scale=0.68]{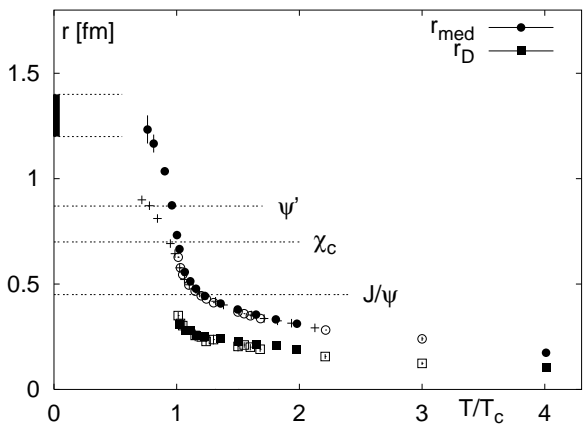}
\vskip -0.5 cm
\caption{ Debye radii for charmonia vs.
temperature\cite{KacZan}.}  
\vskip -0.5 cm
\label{fig:nine}
\end{figure}

\begin{figure}[htb]
\includegraphics[scale=0.68]{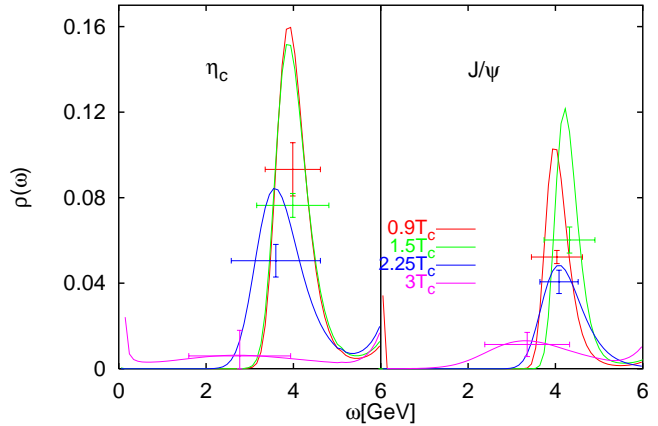}
\vskip -0.5 cm
\caption{ Spectral function of $\eta_c$ and $J/\psi$. From \cite{Saumen}.}  
\vskip -0.5 cm
\label{fig:ten}
\end{figure}

Debye screening of coloured heavy quarks in the deconfined phase had long been
recognised \cite{MatSat} as a possible signal of formation of quark-gluon
plasma, detectable in the suppression of heavy quarkonia in the heavy ion
collisions.  In view of the impressive data from CERN at lower SPS energies,
and the expectations from the upcoming LHC experiments, a critical assessment
of the original theoretical argument seems prudent.  Lattice QCD has
contributed handsomely in finite temperature investigations of both the heavy
quark-antiquark potential, which can be used in the Schr\"odinger equation to
look for the melting of heavy quarkonia, and directly in the spectral function
at finite temperature.  Figure \ref{fig:nine} displays the results
\cite{KacZan} for the screening radii estimated from the inverse
non-perturbative Debye mass $m_D$ in quenched (open squares) and full (filled
squares) QCD.  For $r < r_{med}$, the medium effects are suppressed, leading to
the same heavy quark potential as at $T=0$.  The horizontal lines correspond to
the mean squared charge radii of $J/\psi$, $\chi_c$ and $\psi\prime$ charmonia,
and are thus the averaged separations $r$ entering the effective potential in
potential model calculations.  Figure \ref{fig:nine} therefore suggests that
the $\chi_c$ and $\psi\prime$ states would melt just above the transition while
$J/\psi$ may need higher temperatures to be so affected.  Direct spectral
function calculations \cite{Saumen} provide a strong support for such a
qualitative picture. Such computations have been made feasible by the
recognition of the maximum entropy method (MEM) technique as a tool to extract
spectral functions from the temporal correlators computed on the Euclidean
lattice.  However, as in the case of shear viscosity above, the data for such
temporal correlators are sparse, making the extraction more of an art.
Nevertheless, large lattices, $48^3 \times 12$ to $64^3 \times 24$ have been
used in this case to avoid such criticisms. Figure \ref{fig:ten} shows typical
results for the $J/\psi$ and $\eta_c$ mesons in the quenched approximation.
The vertical error bars denote the possible uncertainties on the area under the
peak as defined by the horizontal error bar.  The peaks in both spectral
functions appear to persist up to 2.25 $T_c$, i.e., have nonzero area within the
computed error-band, and are gone by 3$T_c$ unlike the $\chi_c$ which has no
peak already by 1.1 $T_c$.  Further technical improvements, such as the
inclusion of light dynamical quarks,are clearly desirable.  Another important
issue is that of the huge widths of the peak compared to their known zero
temperature values.  If real, they could hint at rather loosely bound states
which could be dissociated by thermal scatterings.

\subsection{QCD Phase Diagram}

The quark-gluon plasma phase and the corresponding quark-hadron transition
which we discussed so far is a special case of the conditions that could be
created in the heavy ion collisions. Indeed, the lattice QCD thermodynamics
that we considered was for the case of zero net baryon density and an
almost baryon-free region can be produced in the heavy ion collisions in the
so-called central rapidity region, as we explain in the next section.  It
also pervaded our Universe a few microseconds after the Big Bang.  In
general, of course, one should expect hot regions with some baryon number
since the colliding nuclei themselves carry substantial baryon number.
Massive stars could also have regions of huge baryon densities in the core
which could even be at rather low temperatures.  It is natural to ask
what these generalized extreme conditions lead us to.   One could have new
phases, and different natures of phase transitions which may even have
astrophysical consequences.  The vast research area of QCD phase diagram in the
plane of temperature $T$ and the baryonic chemical potential $\mu_B$ deals
with these and several  other interesting issues.  While the current theoretical
expectations suggest such physics at nontrivial baryon densities to be
better accessible to the colliders at lower energies, such at the RHIC in
New York or the forthcoming  FAIR facility at GSI, Darmstadt, we feel that
the physics may be interesting in its own right to be included in this
article dedicated to LHC; with some luck LHC experiments may have important
contributions to this area as well.

Using simple effective QCD models, such as the Nambu-Jana Lasinio model at
finite temperature and densities \cite{buballa}, several speculations have
been made about how the QCD phase diagram in the $T$-$\mu_B$ plane should
be.  At asymptotically high densities, one expects quarks to be effectively
free, and therefore to exhibit various colour superconducting phases
\cite{RajWil}.   In the limit of large number of colours $N_c$ for quarks,
it has also been argued that a ``quarkyonic'' phase may exist \cite{McPi}
at low enough temperatures.  A crucial question, especially in the context
of either the massive stars, or heavy ion collisions, is the quantitative
reliability of the predicted regions in the $T$-$\mu_B$ space.
Alternatively, it is unclear how low can the asymptotic predictions be
trusted.  Nevertheless, most model considerations seem to converge
\cite{RajWil} on the idea of the existence of a critical point in the
$T$-$\mu_B$ plane for the realistic case of 2 light flavours ($m_u = m_d$)
of dynamical quarks with a moderately heavy strange quark.  Establishing it
theoretically and/or experimentally would have huge profound consequences
in our (non-perturbative) understanding of QCD.

Extending the lattice approach to the case of QCD at finite density has
turned out to be a challenging task at both conceptual and computational
level. In principle, it really is straightforward.  One just has to add a
term $\mu_B N_B = \mu_B \bar \psi \gamma_0 \psi $ term to the fermionic
part of the action, hence the Dirac matrix $M$, in eq.(\ref{therm}).  In
order to eliminate certain spurious divergences, even in the free case, some
care is needed \cite{rvg} and the na{\i}ve form above has to be modified.
A big conceptual block has, however, turned up in form of our inability to
define exact chiral invariance in the presence of the chemical potential
\cite{bgs} : both the Overlap and the Domain Wall fermions lose their exact
chiral invariance for any nonzero $\mu$.   The staggered fermions do
preserve the chiral invariance for nonzero $\mu$. Furthermore, they are
simpler to handle numerically. Again most of the numerical work has
therefore employed the staggered fermions, although they are plagued with
the difficulties of precise definition of flavour and spin as mentioned
earlier.  Indeed, the existence of the critical point depends \cite{RajWil}
crucially on how many flavours of light quarks the theory has.  Proceeding
none the less with the staggered quarks, another tough problem arises in form
of the fact that the Det $M(\mu \ne 0)$ in eq. (\ref{therm}) is complex
whereas the numerical methods of evaluation, employed to obtain the results
in the sections above, work only if the determinant is positive definite.
This is akin to the sign problem well known to the statistical physicists
and is largely unsolved in its full generality. 

\begin{figure}[htb]
\vskip -0.5 cm
\includegraphics[scale=0.65]{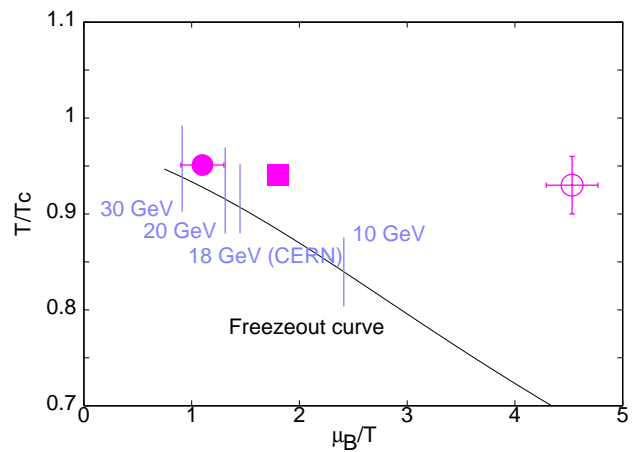}
\vskip -0.5 cm
\caption{ QCD Phase diagram for 2 light flavours of quarks. The circles
\cite{FoKa,gg4} and the square \cite{gg6} denote the location of the critical 
point on lattices with $1/4T$ and $1/6T$ cut-offs respectively. Taken
from \cite{gg4}, where more details can be found.}  
\label{fig:elf}
\end{figure}

A bold breakthrough was achieved \cite{FoKa} by applying the method of
re-weighting in the vicinity of the finite temperature transition at $\mu =
0$.  A flurry of activity saw many new methods emerge \cite{mume}, such as
analytic continuation of computations at imaginary chemical potential and
Taylor series expansions of the free energy.  These have been employed to
get a glimpse of whether a critical point does exist, and if yes, what its
location may be.  The field is really in its infancy and unfortunately at
present no consensus amongst the results obtained so far has emerged.
Figure \ref{fig:elf} exhibits the results obtained for the critical point
for the case of two flavours of light quarks with a pion mass
$m_\pi/m_\rho=0.31\pm0.01$, compared to 0.18 in the real world. 
The results \cite{FoKa,gg4} denoted by circles in the Figure \ref{fig:elf} 
are for a lattice cut-off $a= 1/4T$ whereas the square \cite{gg6} denotes 
the first attempt towards the continuum limit by lowering $a$ to $1/6T$.
Large finite volumes have been observed.  The shift in the location
of the open circle in the Figure \ref{fig:elf} was shown\cite{gg4} to be due to the
use of a 10 times larger volume than the open circle \cite{FoKa}. In order
to be brief, we prefer to close this section by noting that different 
results have been claimed in the literature for larger pion masses and for
a different number of flavours. It is hoped that a clear and solid picture
will emerge in the near future.


\section{Relativistic Heavy-Ion Collisions}

At energies of a few GeV/N to a few 10's of GeV/N, colliding nuclei
tend to stop each other thereby forming a dense, baryon-rich matter. At
higher energies, they nearly pass through each other forming a dense,
nearly baryon-number-free matter in the mid-rapidity region. This is
evident in the shapes of rapidity distributions ($dN/dy$ vs $y$) of
the net proton (i.e., proton$-$antiproton) production observed at
various beam energies. This apparent transparency of nuclear matter at
ultra-relativistic energies can be understood in the space-time
picture of the collision, proposed by Bjorken \cite{bj1,bj2}.


\subsection{Bjorken Picture}

Consider, for simplicity, a central (i.e., head-on or zero impact
parameter) collision of two identical spherical nuclei in their CM
frame. Coordinate axes are chosen such that the two nuclei approach
each other along the $z$-axis and collide at the origin at time
$t=0$. Deep inelastic scattering experiments have revealed the parton
structure of hadrons: In the proton, e.g., the valence quark
distributions $xu_v(x), ~xd_v(x)$ peak around $x \sim 0.2$  and
vanish as $x \rightarrow 0/1$. ($x$ is the Bjorken scaling variable.)
The gluon and sea quark distributions, $xg(x), ~xu_s(x), ~xd_s(x)$, on
the other hand, shoot up as $x \rightarrow 0$. These numerous
low-momentum partons are called {\it wee partons}. As a result of the
Lorentz contraction, the longitudinal (i.e., parallel to the beam
axis) spread of the valence quark wave function is reduced to $\sim
2R/\gamma$ where $R$ is the nuclear radius and $\gamma$ its Lorentz
factor. However, no matter how high the beam energy (or $\gamma$), the
incoming nuclei always have in them wee partons with typical momenta
$p \sim \Lambda_{QCD}$, and hence longitudinal spread $\sim 1 $ fm
\cite{bj1}. The wee partons prevent the nucleus from shrinking below
$\sim 1$ fm in the $z$-direction. If $2R/\gamma < 1$ fm, they play an
important role in the collision dynamics.

As a result of the collision of two nuclei, or rather two clouds of
wee partons, a highly excited matter with a large number of virtual
quanta is created in the mid-rapidity region. (In the modern parlance
one talks about coherent ``glasma'' formed by a collision of two
sheets of ``colour glass condensates (CGC)'' \cite{lappi}.)
Hereinafter we discuss only the mid-rapidity region. The virtual
quanta need a finite time ($\tau_{\rm dec}$) to decohere and turn into
real quarks and gluons. Here $\tau_{\rm dec}$ refers to the rest frame
of an individual parton. In the overall CM frame, the relevant time is
$\gamma \tau_{\rm dec}$ due to the time dilation, $\gamma$ being the
Lorentz factor of the parton. It is now clear that ``slow'' partons
decohere earlier and hence near the origin, than the ``fast'' ones
which emerge later at points farther away from the origin. (This is
known as the inside-outside cascade.) In other words, the
large-$x$ part of each nuclear wave function continues to move along
its light-cone trajectory leaving the small-$x$ part behind. Thus, in
the limit of high beam energy, the time dilation effect causes the
near transparency of nuclei, referred to earlier.

\begin{figure}[htb]
\begin{center}
\includegraphics[scale=0.45]{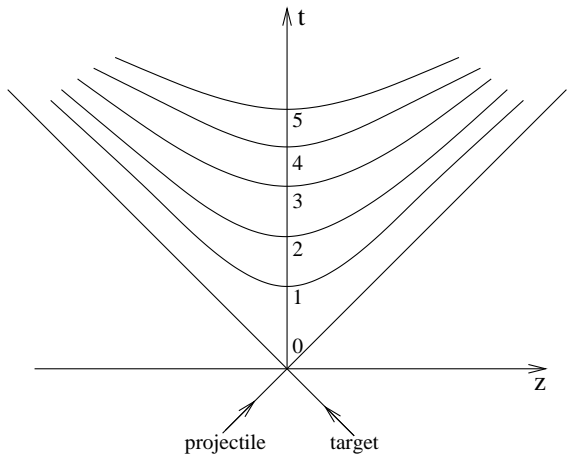}
\vskip -0.5 cm
\caption{Space-time picture of an ultra-relativistic nucleus-nucleus
collision in $1+1$ D for simplicity}
\vskip -0.5 cm
\label{fig:bj}
\end{center}
\end{figure}

Figure \ref{fig:bj} shows this schematically in $1+1$ dimension for
simplicity. The curves are hyperbolas of constant proper time $\tau =
\sqrt{t^2-z^2}$. All points on a given hyperbola are at the same stage
of evolution. In particular, let the hyperbola labelled `1' refer to
$\tau=\tau_{\rm dec}$. Parton at $z$ undergoes decoherence at time
$t=\sqrt{\tau_{\rm dec}^2+z^2}$. The larger the $z$, the larger the
time $t$ and higher the parton velocity $v_z=z/t$ \cite{bj2}.

If the partons thus formed interact amongst themselves a multiple
number of times, the system approaches local thermal equilibrium.
Thermalization time $\tau_{\rm th} ~(> \tau_{\rm dec}$) is estimated
to be of the order of 1 fm.

Figure \ref{fig:bj} indicates a possible scenario. $1,...,5$ are the
hyperbolas with proper times $\tau_1, ..., \tau_5$.\\
$t=0=z$ : the instant of collision\\
$0<\tau<\tau_1$ : formation of quark-gluon matter\\
$\tau_1<\tau<\tau_2$ : (local) equilibration of quark-gluon \\
\hspace*{2cm} matter, i.e., formation of QGP \\
$\tau_2<\tau<\tau_3$ : hydrodynamic evolution of QGP \\
\hspace*{2cm} (partonic EOS)\\
$\tau= \tau_3$ : hadronization\\
$\tau_3<\tau<\tau_4$ : hydrodynamic evolution (hadronic EOS) \\
$\tau_4<\tau<\tau_5$ : transport theoretic evolution of hadrons \\
$\tau= \tau_5$ : freezeout \\
$\tau> \tau_5$ : free-streaming to detectors \\

The above is a rather simple-minded picture: in reality, there are no
such ``water-tight compartments''. The framework of hydrodynamics is
applicable, if at all, only when the system is at or near (local)
thermal equilibrium. If the matter formed in ultrarelativistic
heavy-ion collisions is fully thermalized, one may use the framework
of relativistic ideal fluid dynamics to study its evolution. If it is
only partially thermalized, one could use relativistic dissipative
fluid dynamics. In any case, the covariant transport theory provides a
more general framework for this purpose.

Bjorken \cite{bj2} presented the following formula to estimate the
energy density attained in the mid-rapidity region:
\beq
\varepsilon_0=\frac{1}{\pi R^2 \tau_f}\frac{dE_T}{dy}
~ {\rule[-.28cm]{.01mm}{.8cm}}_{~y=0},
\label{bjform}
\eeq
where $R$ is the nuclear radius, $\tau_f \sim 1$ fm/c is the formation
time of QGP, and $E_T$ is the transverse energy.

It is clear that even if QGP is formed, its lifetime will be of the
order of a few fm/c or $\mathcal{O}(10^{-23})$ seconds, and what
experimentalists detect in their detectors are not quarks or gluons,
but the standard hadrons, leptons, photons, etc. It is a highly
nontrivial task to deduce the formation of QGP from the properties of
the detected particles. This is analogous to the situation in
cosmology where one tries to deduce the information on the early
epochs after the Big Bang by studying the cosmic microwave background
radiation and its anisotropy.

Actually the analogy between the Big Bang and the ``Little Bang'' is
quite striking. In both the cases the initial conditions are not
accurately known, but there are plausible scenarios. In the former
case, there is inflation occurring at $\sim 10^{-35}$ sec, with the
inflaton energy converting into matter and radiation, leading to a
thermal era. In the latter case, one talks about a highly excited but
coherent glasma converting, on the time scale of $\sim 10^{-24}$ sec,
into quarks and gluons which may thermalize to form QGP. In both the
cases the ``fireball'' expands, cools, and undergoes one or more (phase)
transitions. Decoupling or freezeout follows --- of photons in the
former case and of hadrons in the latter. The unknown initial
conditions are parameterized and one tries to learn about them by
working one's way backwards, starting with the detected particles. As
we shall see shortly, the anisotropy of the detected particles plays a
crucial role in the diagnostics of the Little Bang too.

{\it Definition}: The STAR collaboration at RHIC has defined the QGP
as ``a (locally) thermally equilibrated state of matter in which
quarks and gluons are deconfined from hadrons, so that colour degrees
of freedom become manifest over {\it nuclear}, rather than merely
nucleonic, volumes'' \cite{starwp}. The two essential ingredients of
this definition are (a) local equilibration of matter, and (b)
deconfinement of colour over nuclear volumes. Recent claims of the
discovery of QGP at RHIC \cite{gyu} were based on two observations
which, for the first time, provided a good evidence that each of these
two requirements has been fulfilled. We discuss them one by one in the
next two subsections ({\bf 3.2, 3.3}). That will be followed by brief
descriptions of a few other signals of QGP in subsections {\bf 3.4,
3.5}.


\subsection{Anisotropic Flow}

\begin{figure}[htb]
\begin{center}
\includegraphics[scale=0.40]{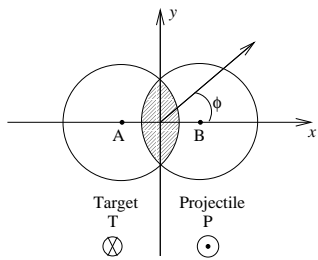}
\vskip -0.5 cm
\caption{Non-central collision}
\vskip -0.5 cm
\label{fig:colli1}
\end{center}
\end{figure}

\begin{figure}[htb]
\begin{center}
\includegraphics[scale=0.55]{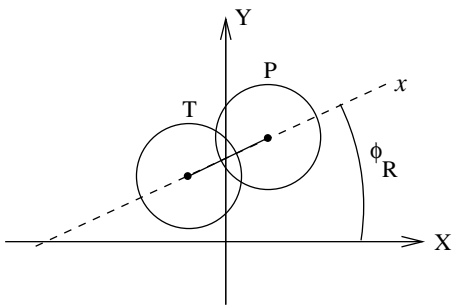}
\vskip -0.5 cm
\caption{Non-central collision. XY are lab-fixed axes.}
\vskip -0.5 cm
\label{fig:colli2}
\end{center}
\end{figure}

Consider now a non-central (or non-zero impact parameter) collision of
two identical (spherical) nuclei travelling in opposite directions.
Choose $x,y$ axes as shown in Fig. \ref{fig:colli1}. The collision or
beam axis is perpendicular to the plane of the figure. Length of the
line AB connecting the centres of the two nuclei is the impact
parameter $b$. Plane $xy$ is the azimuthal or transverse plane. Plane
$xz$ is the reaction plane. It is determined by the impact parameter
vector $\bf b$ and the collision axis. (Obviously the reaction plane
cannot be defined for a central collision.) $\phi=\tan^{-1}(p_y/p_x)$
is the azimuthal angle of an outgoing particle. The almond-shaped
shaded area is the overlap zone. In a real experiment,
Fig. \ref{fig:colli2}, the $x,y$ axes need not coincide with the
lab-fixed $X,Y$ axes. Indeed the reaction plane subtends an arbitrary
angle $\phi_R$ with the $X$ axis. $\phi_R$ varies from event to event.
It is a priori unknown and special experimental techniques are needed
for its determination.


The triple differential invariant distribution of particles emitted in
the final state of a heavy-ion collision is a periodic even function
of $\phi$, and can be Fourier decomposed as
\begin{eqnarray}
E\frac{d^3N}{d^3p}&=&\frac{d^3N}{p_Tdp_Tdyd\phi} \nonumber\\
&=&\frac{d^2N}{p_Tdp_Tdy}\frac{1}{2\pi} \left[ 1+ \sum_1^\infty
2 v_n \cos(n\phi) \right] \nonumber,
\end{eqnarray}
where $y$ is the rapidity and $\phi$ is measured with respect to the
reaction plane. The leading term in the square brackets in the above
expression represents the azimuthally symmetric {\it radial
flow}. $v_1$ is called the {\it directed flow} and $v_2$ the {\it
elliptic flow}. $v_n \equiv \mean{\cos(n\phi)}$ is actually a function of
$p_T$ and $y$. Here the average is taken with a weight equal to the
triple differential distribution of particles in the $(p_T,y)$ bin
under consideration.
$v_2$ can also be written as $\mean{(p_x^2-p_y^2)/(p_x^2+p_y^2)}$.
For a central collision the distribution is
azimuthally isotropic and hence $v_n=0$ for $n=1,2,...$. In other
words, only the radial flow survives.

Measurement of the radial flow: Radial flow gives a radially outward
kick to the emerging hadrons thereby depleting the low-$p_T$
population and making their $p_T$ spectra flatter. The heavier the
hadron, the stronger the momentum kick it receives. By measuring the
slopes of the $p_T$ spectra of various hadrons, the radial flow
velocity can be extracted. At RHIC it turns out to be a sizeable
fraction ($\sim 50$\%) of the speed of light. Thus the flow is
compressible.

Measurement of the anisotropic flow $v_n$: There are several
methods. (a) The most obvious one is based
on the definition $v_n \equiv \mean{\cos n(\phi-\phi_R)}$ where both
$\phi$ and $\phi_R$ are measured with respect to a lab-fixed frame of
reference. This, however, requires the knowledge of $\phi_R$ which
varies from event to event and is not easy to determine. (b)
Two-particle correlation method: This gives $v_n^2=\mean{\cos n
(\phi_1-\phi_2)}$, where $\phi_1$ and $\phi_2$ are azimuthal angles of
two outgoing particles. This method has an advantage that the reaction
plane need not be known. However, $v_n$ is determined only up to the
sign. There are several other methods such as the cumulant method
\cite{borghini}, mixed-harmonic method \cite{mixedh}, Lee-Yang zeroes
method \cite{rsb}, etc. For a recent review, see \cite{volo}.

Importance of the anisotropic flow $v_n$: Consider a non-central
collision, Fig. \ref{fig:colli1}. Thus the initial state is
characterized by a spatial anisotropy in the azimuthal
plane. Consider particles in the almond-shaped overlap zone. Their
initial momenta are predominantly longitudinal. Transverse momenta, if
any, are distributed isotropically. Hence $v_n ({\rm initial}) =
0$. Now if these particles do not interact with each other, the final
(azimuthal) distribution too will be isotropic. Hence $v_n ({\rm
final}) = 0$.

On the other hand, if these particles interact with each other a
multiple number of times, then the (local) thermal equilibrium is
likely to be reached. Once that happens, the system can be described
in terms of thermodynamic quantities such as temperature, pressure,
etc. The {\it spatial anisotropy} of the almond-shaped overlap zone
ensures anisotropic pressure gradients in the transverse plane. This
leads to a final state characterized by a {\it momentum anisotropy} in
the $p_x p_y$ plane or equivalently\footnote{Since $\phi = \tan^{-1}
(p_y/p_x)$.} to an anisotropic distribution of particles in the
transverse $(xy)$ plane, and hence a nonvanishing $v_n$. {\it Thus
$v_n$ is a measure of the degree of thermalization of the matter
produced in a noncentral heavy-ion collision.}

\medskip
To sum up, if either of the two ingredients, namely initial spatial
anisotropy and adequate rescatterings, is missing, there is no 
anisotropic flow $(v_n)$.
\medskip

Sensitivity of $v_n$ to properties of matter at {\it early} times
($\sim$ fm/$c$): We saw above that the spatial anisotropy of the
initial state (together with multiple rescatterings) leads to more
matter being transported in the directions of the steepest pressure
gradients, and thus to a non-zero $v_n$. That in turn results in the
reduction in spatial anisotropy (``self-quenching''). In other words,
expansion of the source gradually diminishes its spatial
anisotropy. Thus $v_n$ builds up early (i.e., when the spatial
anisotropy is significant) and tends to saturate as the spatial
anisotropy continues to decrease. (This is unlike the radial flow
which continues to grow until freeze-out and is sensitive to early- as
well as late-time history of the matter). Thus $v_n$ is a measure of
the degree of thermalization of the matter produced {\it early} in the
collision. In other words, $v_n$ is a signature of pressure at {\it
early} times.

Hydrodynamic calculations of $v_n$ involve the equation of state of
QGP. Thus one hopes to learn about the material properties of the
medium, such as the speed of sound, sheer and bulk viscosities,
relaxation times, etc.

Flow may also be affected by the dynamics of the hadronic phase. Study
of the flow would provide constraints on the properties of hadronic
matter too. (It is expected that at LHC, the relative contribution of
the QGP phase to $v_n$ would be larger than that at SPS and RHIC. This
would reduce the effect of the uncertainties in the hadronic phase).

It should, however, be kept in mind that the initial conditions for
the hydrodynamic evolution are not known with certainty. Hence the
task of unravelling the properties of medium is not as easy as it may
appear.

\begin{figure}[htb]
\begin{center}
\includegraphics[scale=0.46]{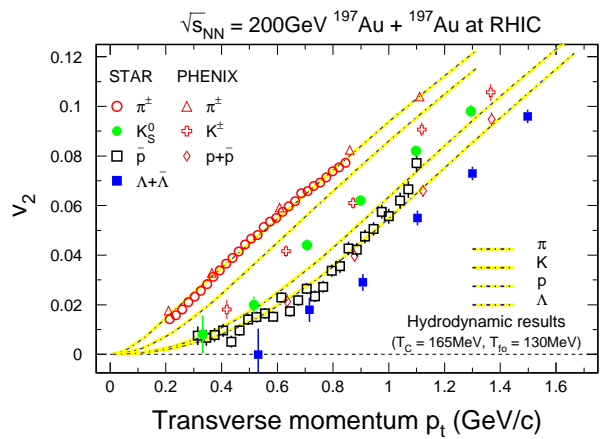}
\vskip -0.5 cm 
\caption{Minimum-bias data. Curves represent ideal hydro results with
a first-order QGP-hadron phase transition. Figure taken from
\cite{STAR}.}
\vskip -0.5 cm
\label{fig:v2pt}
\end{center}
\end{figure}

Figure \ref{fig:v2pt} shows the impressive agreement between RHIC data
on $v_2(p_T)$ and ideal hydro calculations for $p_T$ up to $\sim 1.5$
GeV/c. In particular note the mass ordering: the heavier the hadron,
the smaller the $v_2(p_T)$. This can be understood heuristically as
follows.

Mass ordering of $v_2(p_T)$: Recall that the radial flow depletes the
population of low-$p_T$ hadrons (by shifting them to larger values of
$p_T$). This effect is more pronounced for larger flow velocities and
for heavier hadrons. Suppose $v_2$ is positive as at RHIC, which means
more hadrons emerge in-plane ($x$-direction) than out-of-plane
($y$-direction). Now due to higher pressure gradients in the
$x$-direction, hadrons which emerge in-plane experience a larger flow
velocity than those which emerge out-of-plane. So the depletion is
greater for the hadrons emerging in-plane than out-of-plane. This
tends to reduce the anisotropy and hence $v_2$ of all hadron
species. For a heavier hadron species this reduction is more
pronounced. The net result is $v_2^{light~ hadron} (p_T) > v^{heavy~
hadron}_2 (p_T)$. Mass-ordering signifies a common radial velocity
field.

Hydrodynamic model calculations predicted mass ordering of
$v_2(p_T)$. The broad agreement between the RHIC data and the
predictions of ideal hydro (Fig. \ref{fig:v2pt}) led to the claims of
thermalization of matter and discovery of a perfect fluid --- more
perfect than any seen before.

In order to claim the discovery of a new state of matter, namely
quark-gluon plasma, one needs to demonstrate unambiguously that
(local) equilibrium is attained. There are indications that the
equilibrium attained at RHIC is incomplete \cite{ineqlbm}.

\subsubsection{Constituent Quark Scaling}

\begin{figure}[htb]
\begin{center}
\includegraphics[scale=0.2]{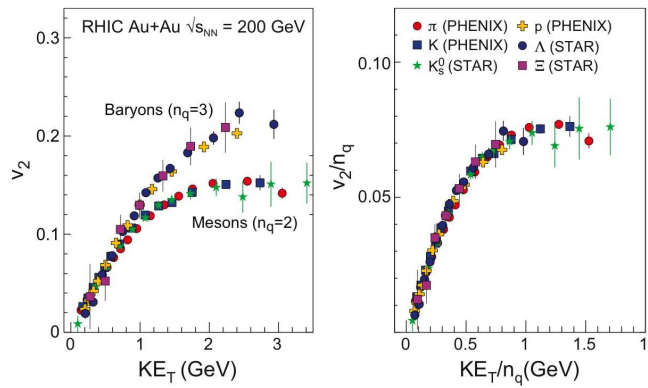}
\vskip -0.5 cm
\caption{Left: Note the two distinct branches. Right: Universal curve. 
Figure taken from \cite{berndt}.}
\vskip -0.5 cm
\label{fig:cqs}
\end{center}
\end{figure}

For $p_T \gtap 2$ GeV/c, ideal hydro results are in gross disagreement
with the $v_2(p_T)$ data: calculated $v_2(p_T)$ continues to rise with
$p_T$, while the data tend to saturate and the mass ordering is reversed.
In the intermediate momentum range (2 GeV/c $\ltap p_T \ltap$ 5 GeV/c),
it is observed that the $v_2/n_q$ vs $p_T/n_q$ (or $KE_T/n_q$) data
fall on a nearly universal curve; see Fig. \ref{fig:cqs}. Here $n_q$
is the number of constituent quarks and $KE_T$ is the transverse
kinetic energy. This is called the constituent quark scaling. It shows
that the flow is developed at the quark level, and that the
hadronization occurs by quark recombination.

\subsection{Jet Quenching}


A variety of signatures of quark-gluon plasma have been proposed.  Some of
the more popular ones are excess strangeness production, thermal dileptons
and photons, jet quenching, $J/\psi$-suppression and event-by-event
fluctuations.  A common theme underlying all of these is the idea of
exploiting the consequences of those properties of QGP which distinguish it
from alternatives like a hot hadron gas.  Since QGP is expected to
form and exist predominantly in the early phase of the collision, the
so-called hard probes are potentially the cleaner direct probes of this
early phase.  It is experimentally known that rare but highly energetic
scatterings produce jets of particles : $ g + g  \to g + g $, where
energetic gluons from the colliding hadrons produce two gluons at large
transverse momenta, which fragment and emerge as jets of showering
particles.  Their typical production time scale is $t \sim 1/Q$, where $Q =
p_T$, the transverse momentum of the jet, is the hard scale of production.
Thus jets at large transverse momenta are produced very early and by
traversing through the produced medium carry its memory while emerging out.
Quark-Gluon Plasma, or any medium in general, interacts with the jet, causing
it to lose energy. This phenomenon goes by the name of jet quenching.

Using the well-known factorization property of perturbative QCD
\cite{qcdbook}, which allows a separation between the hard and soft scales,
a typical cross section at hard scale, say that of hadron $h$ at large
transverse momenta in the process $A + B \to h + X$, can be symbolically
written as 
\begin{eqnarray}
\nonumber
\sigma^{AB \to h} &=& f_A(x_1,Q^2) \otimes f_B(x_2,Q^2) \otimes\\ 
&&\sigma (x_1,x_2, Q^2) \otimes D_{i \to h} (z, Q^2)~.
\end{eqnarray}
\noindent Here $f_A$, $f_B$ are parton distribution functions of the
colliding hadrons $A$ and $B$ at scale $Q^2$, $\sigma (x_1,x_2,Q^2)$ is the
elementary pQCD cross section for partons of momentum fractions $x_1$ and
$x_2$ to produce a parton $i$ with the hard scale $Q = p_T$ for jet
production, and $ D_{i \to h} (z, Q^2)$ is its fragmentation function to
hadron $h$ with momentum fraction $z$.  Various convolution integrations are
denoted symbolically by $\otimes$.  Clearly, there are many more details
which are not spelt out here for brevity,  such as the kinematic
integration region or the summation over all allowed many parton level
processes, such quark-quark or gluon-quark etc.   These can be found in
textbooks \cite{qcdbook}.

\begin{figure}[htb]
\begin{center}
\includegraphics[scale=0.66]{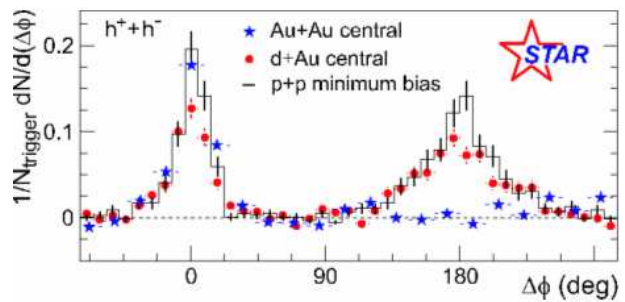}
\vskip -0.5 cm 
\caption{Comparison of the various dihadron angular correlations.  Taken from
\cite{dAustar}.}
\vskip -0.5 cm
\label{fig:dAupt}
\end{center}
\end{figure}

\begin{figure}[htb]
\begin{center}
\includegraphics[scale=0.40]{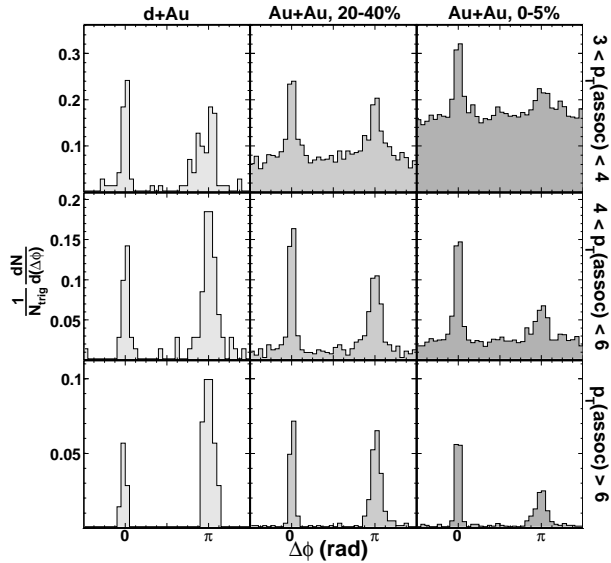}
\vskip -0.5 cm 
\caption{Comparison of the dihadron azimuthal correlations as a function of
the associated $p_T$ for 8 $ < p_T^{trigg} < $ 15 GeV.  Taken from
\cite{higptstar}.}
\vskip -0.5 cm
\label{fig:higpt}
\end{center}
\end{figure}

In presence of a medium, of hot hadron gas or quark-gluon plasma, the
function $D$ above will get modified by the interactions with medium.  The
medium provides scattering centers for the fast moving seed particle of the
jet which typically impart a transverse momentum kick to it.  The medium
induced transverse momentum squared per unit path length, $ \hat q$,
characterizes  the quenching weight function $P(\Delta E)$ \cite{qnwt}
which is the probability that a hard parton loses an additional energy
$\Delta E$ due to its interactions with the medium.  In hot matter with a
temperature of about $T = 250$ MeV, a perturbative estimate \cite{bdmps} for
$ \hat q$ is about 0.5 GeV$^2$/fm.  It is typically a lot smaller in the
cold nuclear matter.   In terms of the quenching weight, one can write
down \cite{qnwt} a medium modified fragmentation function for a jet passing
through a medium as

\begin{equation}
D^{\rm med}_{i \to h} (x, Q^2) = \int_0^1 d \epsilon \frac{P_E(\epsilon)}
{1 -\epsilon}  D_{i \to h}(\frac {x}{1-\epsilon}, Q^2)
~.
\end{equation}
\noindent For a heavy quarkonium like $J/\psi$, the analogue of $D$, is the
wave function of a heavy quark-antiquark pair ($c \bar c$), and it will be
presumably flatter in a hot medium, corresponding to ``its melting''.

RHIC experiments have cleverly exploited their capabilities to perform
tests which have an on-off nature and are therefore rather convincing about
the qualitative existence of the jet quenching phenomenon in heavy ion
collisions. In the case of the elementary  $ g + g  \to g + g $ hard
process, one expects back-to-back jets, i.e, a well-determined azimuthal
correlation between the fast particles.   As jets are hard to identify in
the complex multi-particle environment at RHIC, the STAR collaboration
constructed the angular correlation of hadrons, using a high transverse
momentum $p_T^{trigg}$ particle as the trigger, and studying the azimuthal
distributions of the associated particles ($ p_T^{assoc} < p_T^{trigg}$).
Figure \ref{fig:dAupt} compares the results for gold-gold central
collisions, where one expects formation of a hot medium, with the
proton-proton or deuterium-gold collisions, where one expects to have
turned off the medium effects.   The expected correlation, signalling a
lack of any quenching/medium, is clearly visible in the two peaks separated
by $180^\circ $ for the d-Au and $pp$ collisions.  Remarkably the gold-gold
central collision data show only the peak at zero degree or the near-side.
A hint of the creation of some medium is given by the vanishing of the
away-side jet, at $180^\circ$ degrees, which appears to have been fully
quenched by the medium.  For high enough trigger $p_T$, one can do the same
comparison as a function of range of the associated $p_T$.  Clearly, as the
$p_T^{assoc}$ increases, one ought to see the away-side re-emerge.  This is
beautifully seen in the Figure \ref{fig:higpt}.  It shows the azimuthal
correlations for $8 < p_T^{trigg} < 15 $ GeV for d-Au, and Au-Au collisions
in two centrality bins, with the data for most central collisions displayed
in the last column.  The $p_T$ of the associated particle is restricted to
ranges marked on the right side, and increases as one goes from top to the
bottom.  All panels show comparable strengths for the near-side peak.  As
the $p_T^{assoc}$ grows above 6 GeV, the away-side peaks in all the three
systems also show comparable strengths whereas for lower $p_T^{assoc}$
ranges one has diminishing away-side peaks, characteristic of
jet-quenching.   The same phenomena can also be studied by varying the
$p_T^{trigg}$ and the away-side peak is seen clearly to emerge as
$p_T^{trigg}$ increases. 

\begin{figure}[htb]
\begin{center}
\includegraphics[scale=0.40]{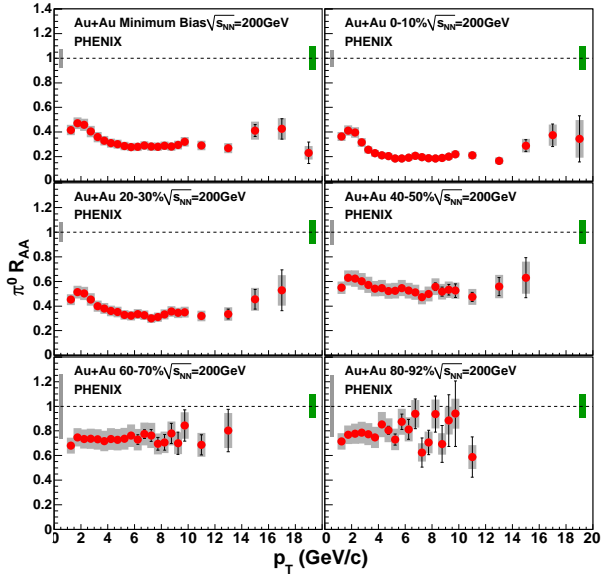}
\vskip -0.5 cm 
\caption{Nuclear modification factor, $R_{AA}$, for neutral pions as a function
of transverse momentum for different centralities.  Taken from
\cite{phenixQ}.}
\vskip -0.5 cm
\label{fig:RAApt}
\end{center}
\end{figure}

A more quantitative investigation of the jet quenching phenomena needs to
extract the transport coefficient $\hat q$, and establish the presence of
the hot matter by comparing it with the corresponding theoretical
estimates, directly from QCD. Many such attempts have been
made.  Recently, the PHENIX experiment \cite{phenixQ} reported their
measurement of neutral pion production in Au-Au collisions at 200 GeV at
the RHIC collider in BNL.   They define the now-famous nuclear suppression
factor $R_{AA}$ as the weighted ratio of the nuclear differential
distribution in rapidity $y$ and transverse momentum $p_T$ and their own earlier
measurements for the same quantity in proton-proton,

\begin{equation}
R_{\rm AA} = \frac{1/{N_{\rm evt}} dN/dydp_T}
{{\langle T_{\rm AB} \rangle } d\sigma_{pp}/dydp_T}~,~
\end{equation}
where further details of determinations of various factors above are given
in \cite{phenixQ}. Their results for $R_{AA}$ are displayed in Figure
\ref{fig:RAApt}.   While the first panel shows the results for their entire
data set, the other panels exhibit data for increasing peripherality of the
collisions (indicated by the increasing range of the percentage label of
each panel), or decreasing centrality.   The error bars indicate the
statistical errors, whereas various systematic errors are shown by the
boxes.  Note that if the nucleus-nucleus collisions were merely scaled
proton-proton ones, one expects $R_{AA} = 1$.  What the data in Figure
\ref{fig:RAApt} indicate, however, is a five-fold suppression that is
essentially constant for 5$<p_T<$20 GeV for the most central bin of 0-10
\%.  The qualitative pattern is the same in all centralities, although the
magnitude of suppression comes down. The highest centrality bin was used to
determine the transport coefficient in the the parton quenching model
\cite{Loizides} to obtain  $\hat q  = 13.2 ^{+2.1}_{-3.2}$ GeV$^2$/fm.
Typically, fits with varying model assumptions do tend to yield a $\hat q$
of  5-15 GeV$^2$/fm.   This order of magnitude or so higher value of the
transport coefficient compared to the expectations from perturbative QCD,
$\sim 0.5$, as mentioned above is an unresolved puzzle.   Nevertheless, the
value hints at a hot medium, presumably even stronger interacting than the
pQCD picture, as the cold matter expectations for $\hat q$ are even more in
disagreement with the experimental determination.  Clearly a lot more needs
to be understood from the data by further delving into the detail
predictions of the models and confronting them with data, as \cite{phenixQ}
attempts to do, in order to establish the nature of hot medium produced as 
that of quark-gluon plasma.


Having discussed the two main observations, anisotropic flow and jet
quenching, which lend support to the claims of discovery of QGP at RHIC, we
now discuss some corroborative evidences which strengthen these claims.
There are also surprises in the RHIC data when compared with the
expectations from the earlier lower energy heavy ion collisions at SPS in
CERN.  We discuss some with the aim to prepare ourselves for the
expectation at yet higher energy in LHC.


\subsection{Anomalous $J/\psi$ Suppression}

Amongst the many signatures proposed to look for QGP experimentally, the
idea of $J/\psi$-suppression has attracted the most attention as the likely
``gold-plated'' signal.   Soon after the pioneering work of Matsui and Satz
\cite{MatSat}, arguing that i) as a hard QCD process, the heavy charm pair
production takes place very early, ii) the Debye screening of the QGP
prevents formation of a $J/\psi$ state in heavy ion collisions, and iii)
the low temperatures at the hadronization do not permit production of
charm-anticharm pair kinematically, it was further proposed that
the suppression
pattern ought to have a characteristic \cite{KarPet} transverse momentum
dependence.  Recognising that the gluon and quark distribution functions
depend on the atomic number $A$, known by the famous EMC-effect, it was
shown in a perturbative QCD calculation that the suppression signal
\cite{ggpsi} itself as well as its $p_T$-dependence \cite{ggspsi} can be
mimicked by the mundane nuclear shadowing.  Thus it became clear since the
early days that a detailed quantitative analysis is necessary to disentangle
the effects of the Debye screening in QGP.  It has since been recognised
that other effects, notably the absorption \cite{GerHuf} of the produced
$J/\psi$ in the nucleus, causes suppression of $J/\psi$ in all $pA$ and
$AB$-collisions.  Thus one has to first account for this expected or normal
suppression and then look for additional or anomalous $J/\psi$-suppression as the
possible signal of QGP.  Considering the general wisdom that
$J/\psi$-production can be computed in pQCD, it ought to be a
straightforward task to compute this normal suppression.   Unfortunately, it
is not so.   One reason is that the gluon distribution function, and the
nuclear shadowing effects, are not well known.   Another, perhaps much more
important reason, is that the hadroproduction of $J/\psi$ needs to tackle
the vexing issue of its formation from the perturbatively produced charm-anticharm
pair.  One usually depends \cite{hardpr} on models, such as the colour
evaporation or the color octet model, hoping that the effective theory
descriptions are valid.  It turns out to be true for large $p_T$ charmonium
production but not for the total cross sections of interest for the QGP
signal.
\begin{figure}[htb]
\begin{center}
\includegraphics[scale=0.4]{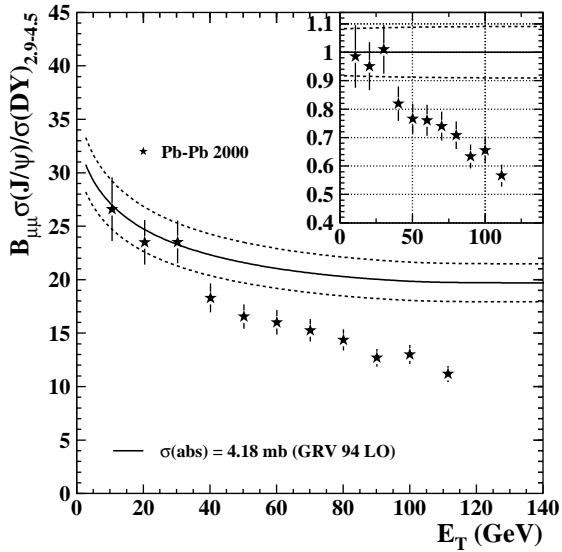}
\vskip -0.5 cm
\caption{$J/\psi$-suppression in Pb-Pb collisions at SPS as a function of
transverse energy $E_T$.  Figure taken from \cite{na50_5}.}
\vskip -0.5 cm
\label{fig:psi1}
\end{center}
\end{figure}

\begin{figure}[htb]
\begin{center}
\includegraphics[scale=0.4]{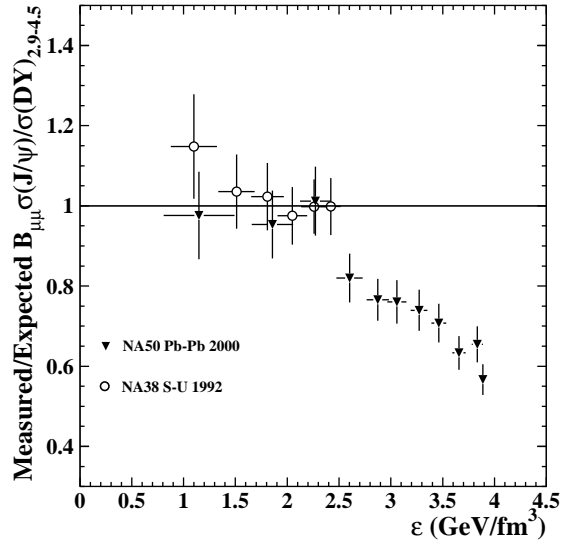}
\vskip -0.5 cm
\caption{$J/\psi$-suppression in Pb-Pb collisions at SPS as a function of
the energy density $\epsilon$.  Figure taken from \cite{na50_5}.}
\vskip -0.5 cm
\label{fig:psi2}
\end{center}
\end{figure}

The preferred phenomenological method \cite{GerHuf} has been to parametrise
the ratio of $J/\psi$-cross sections, total or appropriate differential
cross sections in its transverse momentum $p_T$, or forward momentum
fraction $x_F$ etc., in $pA$ and $pp$ collisions at the same colliding
energy, $\sqrt{s}$, as $\exp (- \sigma_{\rm abs}(J/\psi)\rho_0L)$, where
$L$ is the mean length of the trajectory of the produced $c\bar c$ pair in
nuclear matter and $\rho_0$ is the nuclear density. The parameter,
$\sigma_{\rm abs}(J/\psi)$, is obtained by fitting the data.  Defining a
mean free path $\lambda = 1/ \sigma_{\rm abs}(J/\psi)\rho_0$, one then
extends this idea to the heavy-ion collisions to define the normal or
expected $J/\psi$ suppression due to the traversing of the $c \bar c$-pair
in the nuclear matter as  $\exp(- (L_A+L_B)/\lambda)$.  Here $L_A$ and
$L_B$ are the lengths for the trajectories of the $c\bar c$ in the
projectile ($A$) and target ($B$) respectively. They are calculated from
collision geometry by using the oft-used relations  between mean transverse
energy of the bin, $E_T$, and the average impact parameter $b$. 

Figure \ref{fig:psi1} exhibits \cite{na50_5} the results of the NA50
collaboration on $J/\psi$ cross section as a function of the transverse
energy $E_T$ in Pb-Pb collisions at $\sqrt{s} \simeq 17$ GeV.  It is
normalized to the Drell-Yan cross section in the mass range shown and
$B_{\mu \mu}$ is the branching fraction of $J/\psi$ in the dimuon channel.
The full curve depicts the expected normal suppression as a function of
$E_T$, computed as explained above using the fitted $J/\psi$ cross section
of 4.18 mb obtained from the NA50's own $pA$ data.  The dashed lines show
the computed error bars on the expected suppression, and the inset shows
the ratio of measured to the expected suppression.  Using the Bjorken
formula in eq.  (\ref{bjform}), one obtains this ratio of the measured to
the expected cross section ratio of the $J/\psi$ and the Drell-Yan as a
function of the energy density in GeV/fm$^3$ units, as shown in Figure
\ref{fig:psi2}, taken from \cite{na50_5}.   One sees that the anomalous
suppression, i.e., depletion of the measured cross section from that
expected, sets in at an energy density of about 2.5 GeV/fm$^3$, comparable
to the expectations from lattice QCD, as seen in Figure \ref{fig:four}.   A
natural explanation of the anomalous suppression was, therefore, the
formation of quark-gluon plasma.  Since the $J/\psi$-production takes place
both directly and through other charmonium states like $\chi_c$, the slow
fall-off with the energy density in Figure \ref{fig:psi2} could be
interpreted as gradual progress towards the full suppression.  However, one
could also explain the anomalous suppression in alternative ways, using
hadronic \cite{Capella} or thermal \cite{Gaz} models.  Since one expects
the higher collision energy at RHIC to produce higher temperatures/energy
densities, one expected a further stronger suppression at RHIC. Indeed,
this seems to be true both in the quark-gluon plasma models as well as the
alternatives, the difference between them being quantitative in nature.

\begin{figure}[htb]
\begin{center}
\includegraphics[scale=0.4]{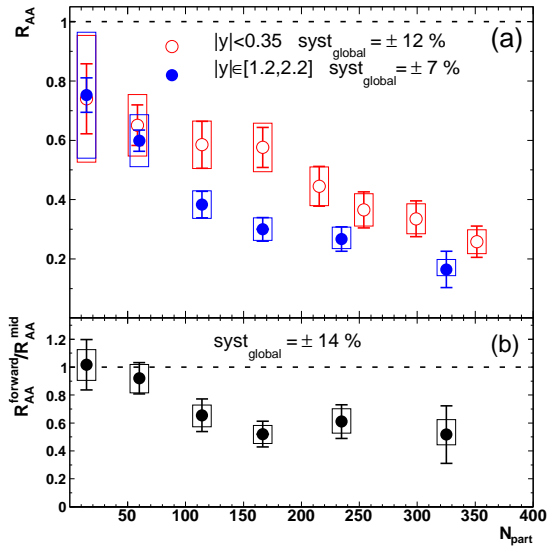}
\vskip -0.5 cm
\caption{$J/\psi$-suppression in Au-Au collisions at PHENIX, BNL as a
function of number of participants.  Figure taken from \cite{phenixPRL}.}
\vskip -0.5 cm
\label{fig:psi4}
\end{center}
\end{figure}
\begin{figure}[htb]
\begin{center}
\includegraphics[scale=0.5]{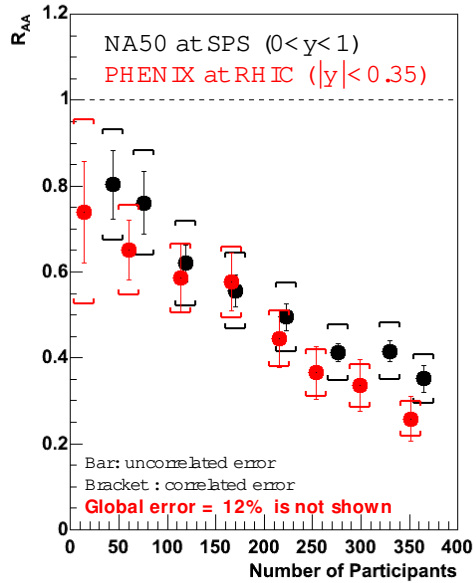}
\vskip -0.5 cm
\caption{Comparison of NA50 and PHENIX results on $J/\psi$-suppression as a
function of number of participants.   Figure taken from \cite{mjt}.} \vskip
-0.5 cm
\label{fig:psi5}
\end{center}
\end{figure}

The RHIC results \cite{phenixPRL}, however, brought a big surprise by being
different from any of those expectations. Analogous to the case of jet
quenching in the previous section, the PHENIX collaboration at RHIC
constructs the ratio $R_{AA}$ of the $J/\psi$ (differential) production
cross section in $AA$ collisions and the corresponding $pp$ cross section
weighted by the number of binary collisions.  Figure \ref{fig:psi4}
displays their results for $R_{AA}$ in Au-Au collisions at $\sqrt{s} =200$
GeV.  They show {\em more} suppression in the forward region ($ |y| \in
[1.2,2.2]$, filled circles in the top panel), than the central ($ |y|  <
0.35$, open circles in the top panel) for number of participants greater
than 100 (alternatively for large enough transverse energy $E_T$).  More
importantly, a direct comparison \cite{mjt} in Figure \ref{fig:psi5}
clearly demonstrates that the PHENIX data in the central rapidity region
are in very good agreement with the CERN NA50 results \cite{na50_5}.  The
trends for both the central region of the CERN and RHIC experiments, as
seen in Figure \ref{fig:psi5}, and the ratio of forward to the central
rapidity region, as seen in the bottom panel of Figure \ref{fig:psi4}, are
against \cite{mjt} the predictions of the models which successfully
accounted for the NA50 data.

There have been some attempts to solve this $J/\psi$-puzzle.  As we saw in
the Figure \ref{fig:ten} of section {\bf 2.4}, the lattice QCD results
suggest melting of the $J/\psi$ takes place at higher temperatures ($ > 2
T_c$) than predicted by simple models.  A way to understand the results in
Figure \ref{fig:psi5} could then suggest itself if the temperature reached
at both the SPS and RHIC energy is $\ltap 2 T_c$. In that case, only
$\chi_c$ and $\psi'$ would have melted \cite{satz}, suppressing the
corresponding decay $J/\psi$'s, and giving similar results for CERN and
RHIC experiments.    Since the temperature reached at LHC is expected to
cross $2 T_c$, a clear prediction of such a scenario would then be much
more suppression for LHC than that in Figure \ref{fig:psi5}.  However,
there are other scenarios, including thermal enhancement \cite{TheEnh}
arising due to recombination of the large number of thermal produced
charm-anticharm quarks.  These would predict an overall enhancement.
In any case, $J/\psi$-suppression could provide a lot of excitement again
at LHC.


\subsection{Particle ratios \& Bulk Properties}

\begin{figure}[htb]
\begin{center}
\includegraphics[scale=0.45]{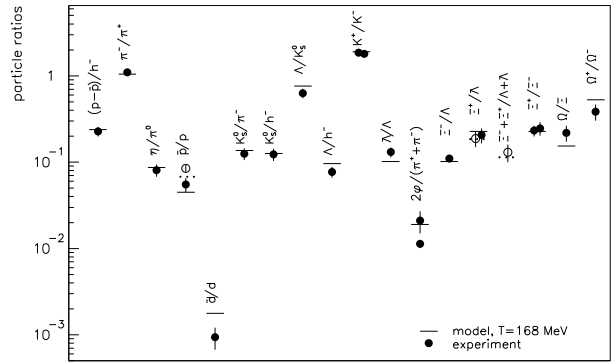}
\vskip -0.5 cm
\caption{Comparison between the statistical model (horizontal bars)
and experimental particle ratios (filled circles) measured at SPS
CERN. From Braun-Munzinger et al. \cite{statmod}.}
\vskip -0.5 cm
\label{fig:ratios}
\end{center}
\end{figure}

A variety of hadrons are produced in an ultra-relativistic heavy-ion
collision. They are identified and their relative yields measured; see
Fig. \ref{fig:ratios}. These hadron abundance ratios can be calculated
in a simple statistical model \cite{statmod}: It is assumed that these
particles emerge from a chemically equilibrated hadron gas
characterized by a chemical potential ($\mu_i$) for each hadron species
and a common temperature ($T$). The number density $n_i$ of hadron of
type $i$ is then given by the standard Fermi-Dirac ($+$) or
Bose-Einstein ($-$) formulas
\[
n_i=d_i \int \frac{d^3p}{(2 \pi)^3}
\frac{1}{\exp\left[(E_i-\mu_i)/T\right] \pm 1} ,
\]
where $d_i$ is the spin degeneracy. At chemical equilibrium, the
chemical potential $\mu_i$ can be written as $\mu_i=\mu_B B_i - \mu_S
S_i -\mu_I I^{(3)}_i$ where $B_i,~S_i$ and $I^{(3)}_i$ stand for the
baryon number, the strangeness and the third component of the isospin
quantum numbers, respectively, of the hadron of type $i$. The two
unknown parameters $T$ and $\mu_B$ are fitted to the data. This simple
model has been quite successful in explaining the SPS and RHIC data;
see Fig. \ref{fig:ratios} for SPS and a similar figure in
\cite{bm2004} for RHIC. Note that even the multistrange particles seem
to be consistent with the model. This suggests that they are produced
in a partonic environment rather than in a hadronic one. $T \equiv
T_{\rm ch}$ is the chemical freezeout temperature. The fitted values
are
\begin{eqnarray}
T_{\rm ch}&=&170 ~{\rm MeV}, ~\mu_B~ =~ 270 ~{\rm MeV}, ~{\rm (SPS)}, 
\nonumber \\
T_{\rm ch}&=&176 ~{\rm MeV}, ~\mu_B~ =~ 41 ~{\rm MeV}, 
~{\rm (RHIC~130~GeV)}, \nonumber \\ 
T_{\rm ch}&=&177 ~{\rm MeV}, ~\mu_B~ =~ 29 ~{\rm MeV}, 
~{\rm (RHIC~200~GeV)}. 
\nonumber
\end{eqnarray}
Note the trend of the chemical freezeout point to approach the
temperature axis of the QCD phase diagram as the collision energy is
increased. Data obtained at the AGS and SIS energies are also
consistent with this trend; see Fig. 1.3 in \cite{lhcp1}. 
For more recent fits to the statistical model, see \cite{bm2006}.

\section {Hydrodynamics}

Hydro plays a central role in modelling relativistic heavy-ion
collisions: It is first used for the calculation of the $p_T$ spectra
and the elliptic flow $v_2$. The resultant energy density or
temperature profiles are then used in the calculations of jet
quenching, $J/\psi$ melting, thermal photon and dilepton production,
etc.

Hydrodynamic framework consists of a set of coupled partial
differential equations for energy density, number density, pressure,
hydrodynamic four-velocity, etc. In addition, these equations also
contain various transport coefficients and relaxation times.

Hydro is a very powerful technique because given the initial conditions
and the EOS it predicts the evolution of the matter. Its limitation is
that it is applicable at or near (local) thermodynamic equilibrium
only.


\subsection{A Perfect Fluid?}

How robust is the claim of discovery of a perfect fluid at RHIC, or is
there any need of the viscous hydrodynamics for RHIC? A closer scrutiny
shows that the claim is not really robust, and it is necessary to
do viscous hydro calculations:

\noindent $\bullet$ Agreement between data and ideal hydro is far from
perfect. (Ideal) ``hydro models seem to work for minimum-bias data
but not for centrality-selected $\pi$ and ${\bar p}$ data''
\cite{STAR2}.

\noindent $\bullet$ Initial (and final) conditions for the
hydrodynamic regime are uncertain. It is entirely possible that the
ideal hydro mimics viscous hydro if the initial (and/or final)
conditions are suitably tuned. Most ideal hydro calculations so far
have been done with Glauber-type initial conditions. It has recently
been realized that the CGC-type initial conditions yield higher
eccentricity of the overlap zone \cite{hirano}, and hence higher
$v_2$. To push these results down to agree with data, some viscous
corrections are needed. The same is true with fluctuations in the
initial conditions \cite{soco}. Event-to-event fluctuations in nucleon
positions result in higher eccentricity and hence higher $v_2$
\cite{snell}.

\noindent $\bullet$ Some $v_2$ may build up during the pre-equilibrium
(i.e., pre-hydro) regime. Success of ideal hydro may be due to the
neglect of this contribution to $v_2$ in most calculations \cite{fries}.

\noindent $\bullet$ For realistic light quark masses, the
deconfinement transition is known to be a smooth crossover. However,
it seems that the ideal hydro calculations need a first-order
transition for a best fit to the data \cite{pasi}.

\noindent $\bullet$ The shear viscosity to entropy density ratio
($\eta/s$) may be small in the transition region. But there are
indications that the bulk viscosity to entropy density ratio
($\zeta/s$) may be rising dramatically near $T_c$ \cite{kkt}. If this
result holds, QGP discovered at RHIC cannot be called a perfect fluid.

\begin{figure}[htb]
\begin{center}
\includegraphics[scale=0.36,angle=90]{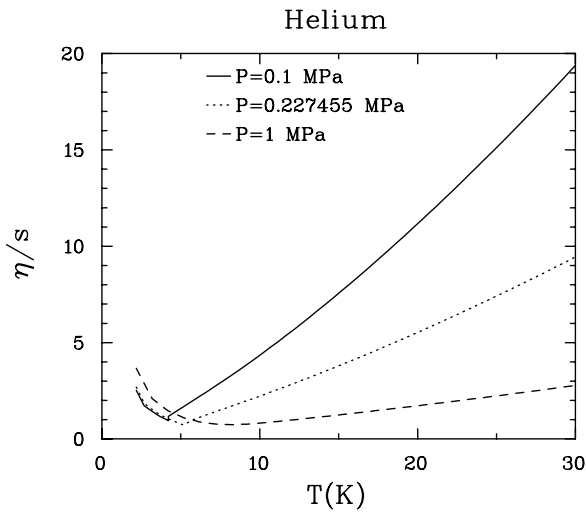}
\vskip -0.5 cm
\caption{
Each curve is at a fixed pressure. Solid: below the critical pressure
$P_c$, dotted: at $P_c$, dashed: above $P_c$. From \cite{csernai}.}
\vskip -0.5 cm
\label{fig:he}
\end{center}
\end{figure}

\noindent $\bullet$ It is known that for helium, water, nitrogen,
$\eta/s$ at constant pressure plotted as a function of temperature,
exhibits a minimum with a cusp-like behaviour at the critical point;
see Fig. \ref{fig:he}. There are indications that the QCD matter too
shows similar trends. Viscous hydro calculations of the QCD matter
would allow us to extract $\eta/s$ from data and might help us
pinpoint the location of the QCD critical point \cite{csernai}.

\noindent $\bullet$ If the inequality $\eta/s > 1/4\pi$ obtained
\cite{kss} from the AdS/CFT duality is applicable to QCD, then also
viscous hydro calculations become necessary.

\noindent $\bullet$ Assume a quasiparticle picture. Quantum mechanical
uncertainty principle tells us that the mean-free path ($\lambda$)
cannot be less than the inverse of the typical momentum of the
quanta. It also makes no sense to have a mean-free path smaller than
the interparticle spacing \cite{dani}. Since $\eta \propto \lambda$,
$\eta$ cannot vanish.

\noindent $\bullet$ Finally, to claim success for ideal hydro, one
should calculate viscous corrections and show explicitly that they are
indeed small.


\subsection{Relativistic Dissipative Hydro --- a Brief History}

Relativistic version of the Navier-Stokes equation was obtained by
Eckart \cite{eckart}, and by Landau and Lifshitz \cite{landau}. This
is called the standard or the first-order formalism because terms only
up to first order in dissipative quantities are retained in the
entropy four-current. (The Euler's equation constitutes the
zeroth-order formalism.) However, it was soon realized that this
formalism suffers from the following problems:

$\bullet$ Acausality: Equations are parabolic and they result in
super-luminal propagation of signals \cite{muller,israel}.

$\bullet$ Instability: Equilibrium states are unstable under small
perturbations for a moving fluid \cite{hiscock}. This makes it
difficult to perform controlled numerical simulations.

$\bullet$ Lack of relativistic covariance: This problem is related to
the previous one. First-order theories look covariant, but they are
not.

A causal dissipative formalism was developed by M\"uller
\cite{muller}, and Israel and Stewart \cite{israel}, in the
non-relativistic and relativistic sectors, respectively. It is also
called a second-order formalism because the entropy four-current now
contains terms up to second order in dissipative quantities. The
resulting hydrodynamic equations are hyperbolic. Application of causal
dissipative hydro to relativistic heavy-ion collisions was pioneered
by Muronga \cite{muronga}. Since then many others have contributed to
this effort. We shall describe some of them in subsection {\bf 4.4}.

Recent years have witnessed intense activity in the area of causal
hydro of gauge theory plasmas from AdS/CFT duality; for reviews see
\cite{ads}.


\subsection{Basic Idea of Causal Dissipative Hydro}

Before we discuss hydrodynamics, let us first consider a simpler
example of diffusion. Consider a fluid in equilibrium with a uniform
density $\rho$. If the fluid is perturbed such that the density is no
longer uniform, it responds by setting up currents which tend to
restore the equilibrium. In the linear response theory, the induced
current $J_i$ is simply proportional to the gradient of $\rho$ (Fick's
law.):
\beq
J_i = -D\partial_i \rho,
\label{ficks}
\eeq
where $D$ is the diffusion coefficient. $D$ is an example of a
transport coefficient. Transport coefficients play an important role
in the study of relaxation phenomena in non-equilibrium statistical
mechanics or fluid dynamics. Equation (\ref{ficks}) connects the
applied force ($-\partial_i \rho$) with the flux ($J_i$). Such
equations are called constitutive equations because they describe a
physical property of the material. (The familiar Ohm's law ${\bf
J}=\sigma {\bf E}$ is another example of this.) In addition to
eq. (\ref{ficks}), we also have the usual current conservation
equation
\beq
\partial_\mu J^\mu = 0.
\label{coneq}
\eeq
If $D$ is constant, elimination of $J_i$ gives
\[
\partial_0 \rho - D \partial_i^2\rho=0.
\]
This is the diffusion equation. It is parabolic. Its
solution is
\[
\rho \sim \exp(-x^2/4Dt) / \sqrt{4 \pi D t}.
\]
It is easy to see that the solution violates causality: 
Initially (i.e., in the limit
$t\rightarrow 0$), this is the Dirac delta function. But at any finite
time, howsoever small, it is nonzero everywhere, even outside the
lightcone. Now eq. (\ref{coneq}) cannot be wrong. So to restore
causality the constitutive equation (\ref{ficks}) which anyway was a
hypothesis, is replaced by
\beq
\tau_J \partial_0 J_i+J_i = -D\partial_i\rho,
\label{ficks2}
\eeq
where $\tau_J$ is a parameter with dimensions of time. In
eq. (\ref{ficks}), if the force vanishes, the flux vanishes
instantaneously without any time lag. In contrast, in
eq. (\ref{ficks2}) the flux relaxes to zero exponentially.  $\tau_J$
is called the relaxation time.  The new diffusion equation is
\[
\tau_J \partial_0^2 \rho + \partial_0 \rho - D \partial_i^2 \rho = 0.
\]
This equation is hyperbolic and is called the Telegraphist's equation
\cite{morse}. If $v^2 \equiv D/\tau_J < 1$, causality is restored.

Now consider hydrodynamics. The conservation and constitutive
equations are
\begin{eqnarray}
\partial_\mu T^{\mu\nu} &=& 0, \nonumber \\
T_{ij} &=& P\delta_{ij}-\eta(\partial_iu_j+\partial_ju_i-\frac{2}{3}
\delta_{ij}\partial_ku_k) \nonumber \\
&-& \zeta \delta_{ij}\partial_ku_k. \nonumber
\end{eqnarray}
Here $T^{\mu\nu}$ is the energy-momentum or stress-energy tensor, $P$
is the equilibrium pressure, and $\eta$ and $\zeta$ are the
coefficients of shear and bulk viscosity, respectively. Tensor
decomposition is now more complicated. But the basic idea remains the
same. Causality is restored by introducing higher-order terms in the
gradient expansion. This forces introduction of a new set of transport
coefficients, e.g., $\tau_\pi$ and $\tau_\Pi$ which are relaxation
times corresponding to shear and bulk viscosities. They are important
at early times or for a rapidly evolving fluid. For details, see
e.g. \cite{muronga}.


\subsection{Recent Results from Causal Viscous Hydro}

The Israel-Stewart formulation \cite{israel} of the causal dissipative
hydro is commonly used for numerical applications. However, it is not
the only causal formulation available. There are others such as
M\"uller's theory \cite{muller}, 
Carter's theory \cite{carter},
\"Ottinger-Grmela formulation \cite{ottinger},
memory function method of Koide et al. \cite{koide}, etc.


We have already mentioned the early work by Muronga \cite{muronga}.
Since then several authors have studied various aspects of the causal
viscous hydro. We now describe briefly only a few of the most recent
of these papers. This will also give the reader a feel for the
complexities of these calculations and the uncertainties therein.
(Other very recent papers which we shall not describe are listed in
\cite{others}.)

\begin{figure}[htb]
\begin{center}
\vskip 0.5 cm
\includegraphics[scale=0.3]{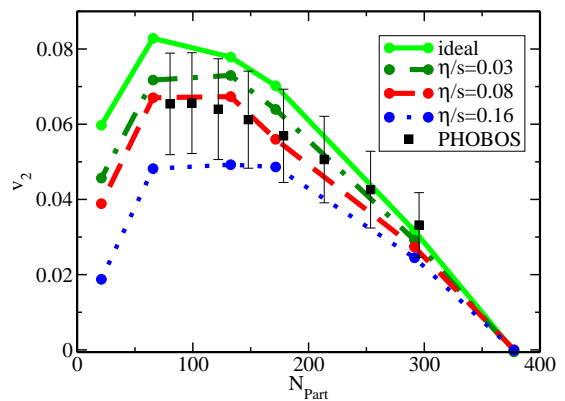}
\vskip -0.5 cm
\caption{Au-Au, 200 GeV, $p_T$-integrated $v_2$ for charged particles
vs number of participant nucleons. PHOBOS: 90\% confidence level
systematic errors. From \cite{roma2}.}
\vskip -0.5 cm
\label{fig:roma1}
\end{center}
\end{figure}

\begin{figure}[htb]
\begin{center}
\includegraphics[scale=0.3]{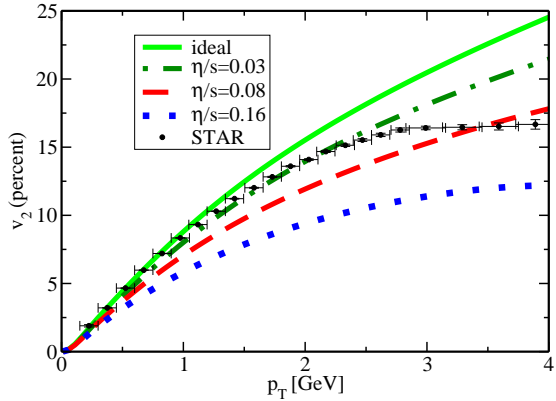}
\vskip -0.5 cm
\caption{Au-Au, 200 GeV, minimum-bias $v_2(p_T)$ for charged
particles. STAR: only statistical errors. From \cite{roma2}.}
\vskip -0.5 cm
\label{fig:roma2}
\end{center}
\end{figure}

Romatschke and Romatschke \cite{roma2} used the Israel-Stewart theory.
They assumed longitudinal boost invariance and used Glauber-type
initial conditions. The initial shear pressure tensor $\pi^{\mu\nu}$
was assumed to be zero. $\eta/s$ was treated as a fixed number
independent of temperature. The bulk viscosity was ignored. For the
EOS they used the semirealistic result of Laine and Schroder
\cite{laine}, and calculated the elliptic flow $v_2$. Their conclusion
was that $p_T$-integrated $v_2$ is consistent with $\eta/s$ up to
0.16; see Fig. \ref{fig:roma1}. However, the minimum-bias $v_2(p_T)$
favoured $\eta/s < 1/4\pi$ violating the KSS bound \cite{kss}; see
Fig. \ref{fig:roma2}.

Dusling and Teaney \cite{dusling} used the \"Ottinger-Grmela formalism
of causal viscous hydro. They assumed longitudinal boost invariance
and used Glauber-type initial conditions. The initial shear pressure
tensor $\pi^{ij}$ was taken to be $\eta \mean{\partial^iu^j}$ as in
the Navier-Stokes theory. $\eta/s$ was treated as a fixed number
independent of temperature. The bulk viscosity was ignored. The EOS
used by them was simply $p=\epsilon/3$ without any phase
transition. Their conclusion was that if the effects of viscosity are
included in the evolution equations but not in the freezeout, then the
$v_2$ is affected only modestly. If, however, they are
included at both the places, then $v_2$ is significantly reduced at
large $p_T$.


Why does the shear viscosity suppress $v_2(p_T)$? Shear viscosity
represents a frictional force proportional to velocity. For an
in-plane elliptic flow, the in-plane flow velocity is higher than that
out of plane. So the in-plane frictional force is stronger. This tends
to reduce the flow anisotropy and hence $v_2(p_T)$.

Calculations described above include the shear viscosity in some
approximation, but ignore the bulk viscosity completely. What do we
know about the bulk viscosity of the strongly interacting matter? In
the high-temperature limit, pQCD calculations \cite{amy} give the
following results for the shear and bulk viscosity coefficients 
\[
\eta \sim \frac{T^3}{\alpha_s^2 \ln \alpha_s^{-1}}~~{\rm and}~~
\zeta \sim \frac{\alpha_s^2 T^3}{\ln \alpha_s^{-1}}.
\]
As $T$ increases, both $\eta$ and $\zeta$ increase. However, the ratio
$\zeta/\eta$ decreases showing the reduced importance of the bulk
viscosity at high $T$. Also note that the entropy density $s \sim
T^3$, and hence $\eta/s$ increases with $T$, whereas $\zeta/s$
decreases with $T$. This is easy to understand because QCD becomes
conformally symmetric at high temperatures.

In the deconfinement transition region the conformal symmetry is badly
broken, and there is no reason to expect the bulk viscosity to be
negligible. Extracting $\zeta$ for temperatures in this region from
lattice QCD is difficult; see section {\bf 2.4}. However, some
preliminary results are now available, and they indicate a dramatic
rise of $\zeta/s$ as $T \rightarrow T_c$ \cite{kkt}.

\begin{figure}[htb]
\begin{center}
\includegraphics[scale=0.6]{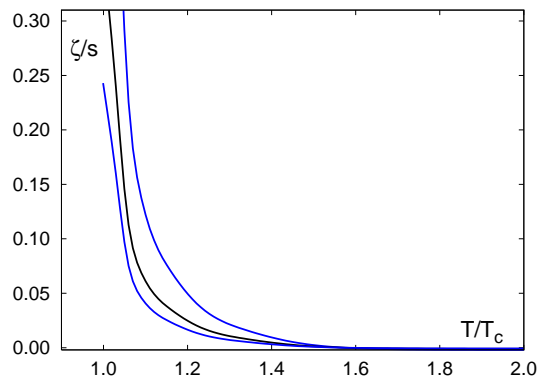}
\vskip -0.5 cm
\caption{Bulk viscosity based on lattice data. $\omega_0=0.5,1,1.5$
GeV (top to bottom) is the scale at which pQCD is applicable. From
\cite{kkt}.}
\vskip -0.5 cm
\label{fig:zeta}
\end{center}
\end{figure}

Taking these results at their face value, Fries et al. \cite{fries2}
have studied the effect of inclusion of the bulk viscosity in the
hydro equations. They studied 1D expansion of the fluid assuming
longitudinal boost invariance. $\eta/s$ was held fixed at $1/4\pi$. A
realistic EOS based on the lattice results of Cheng et
al. \cite{rbrc} was used. Various initial conditions were tried. They
concluded that (a) Large bulk viscosities around $T_c$ lead to
sizeable deviations from equilibrium throughout the entire lifetime of
QGP. (b) Bulk viscosities just slightly larger than currently favoured
could easily lead to breakdown of hydro around $T_c$. (c) The
decreased pressure should slow down the expansion and increase the
time spent by the fluid in the vicinity of the phase transition. (d)
The amount of entropy produced through {\it bulk stress} around $T_c$
is smaller than that produced by {\it shear stress} at earlier
times. Hence no large increase of the final particle multiplicity is
expected.


\subsection{What Remains to be Done?}

\noindent $\bullet$ Bulk as well as shear viscosity (together with
temperature dependence of $\zeta/s$ and $\eta/s$) needs to be
incorporated.

\noindent $\bullet$ Can causal viscous hydro with CGC-type initial
conditions reproduce $dN/dy, \mean{p_T}$ and $v_2$ data? If so, what
are the extracted $\zeta/s, ~\eta/s$?

\noindent $\bullet$ Causal viscous hydro $+$ hadronic cascade is not
done yet.

\noindent $\bullet$ There are issues related to the hydro formalism
itself. For example, Baier et al. \cite{baier} have recently shown
that the M\"uller and Israel-Stewart theories do not contain all
allowed second-order terms.

\noindent $\bullet$ Present uncertainties in the hydro calculations
limit the accuracy with which conclusions can be drawn. A coherent,
sustained collaboration of experts in all stages of heavy-ion
collisions is needed for a detailed, quantitative analysis of
experimental data and theoretical models. Various numerical codes need
to be compared with each other. To that end a new Theory-Experiment
Collaboration for Hot QCD Matter (TECHQM) has been initiated. For
details, see \cite{tech}.


\section{Predictions for LHC}

Pb-Pb collisions at $\sqrt{s_{\rm NN}}=5.5$ TeV is an important part
of the LHC experimental program. 5.5 TeV represents about 30-fold
increase in the CM energy compared to the maximum energy explored at
RHIC which in turn was about 10 times higher than that at
SPS. Measurements on pp collisions as well as collisions of p, d,
light ions with Pb will provide important benchmarks.

Among the experiments at LHC, CMS and ATLAS are primarily particle
physics experiments/detectors, but they will study the physics of
heavy-ion collisions too.  ALICE (A Large Ion Collider Experiment), on
the other hand, is a dedicated heavy-ion collision experiment.
Physicists from several Indian universities and institutions have
contributed in a big way to the ALICE collaboration. They are
responsible for, among other things, the designing, testing,
installation and maintenance of the Photon Multiplicity Detector (PMD)
in ALICE and future upgrades of it. PMD is a preshower detector with
fine granularity, full azimuthal coverage and one unit of
pseudo-rapidity coverage. It will be used to measure the multiplicity,
spatial distribution and correlations of produced photons on an
event-by-event basis. Since photons escape the quark-gluon plasma
without interactions, such measurements can potentially provide a
cleaner glimpse of the early QGP phase. The Indian community has also
made significan contributions to the muon spectrometer of ALICE. The
spectromemter will be useful in the investigations of the $J/\psi$ and
other quarkonia, discussed in subsection 3.4. These particles are
detected via their dimuon decay channel. The muon tracks will be found
with an accuracy of better than one-tenth of a millimeter, thanks to
the state-of-the-art readout electronics, known as MANAS, which was
developed indigenously. ALICE has decided to use a Grid environment
for their computing needs. India is a signatory to the Worldwide LHC
Computing Grid and some of the Department of Atomic Energy
installations are designated as Tier-II centers for this purpose.

A workshop was organized in 2007 at CERN in order to collect all the
existing predictions for heavy-ion collisions at LHC. The proceedings
\cite{armesto} provide a broad overview of the field. Here we shall
only present a few glimpses of what may be in store at LHC. 

\begin{figure}[htb]
\begin{center}
\includegraphics[scale=0.40]{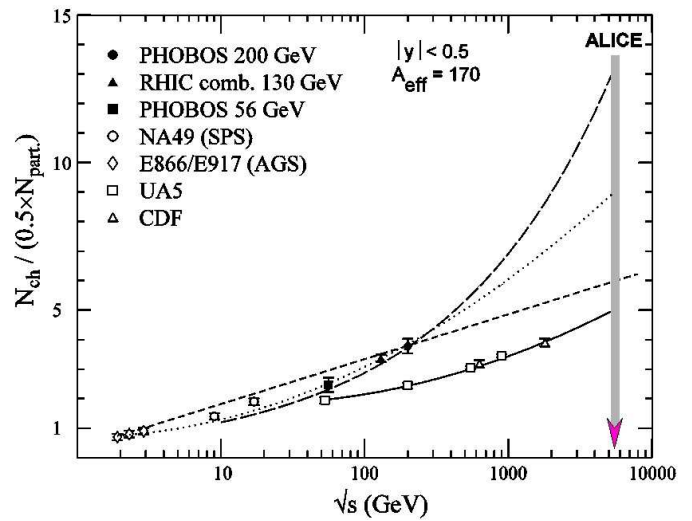}
\vskip -0.5 cm

\caption{Charged-particle rapidity density per participant pair as a
function of center-of-mass energy for AA and pp collisions. Dashed
line: a fit linear in $\ln(\sqrt{s})$, Dotted curve: a fit quadratic
in $\ln(\sqrt{s})$, Long-dashed curve: based on the saturation model
of \cite{ekrt}. From \cite{lhcp1}.}

\vskip -0.5 cm
\label{fig:lhc1}
\end{center}
\end{figure}

\begin{figure}[htb]
\begin{center}
\includegraphics[scale=0.35]{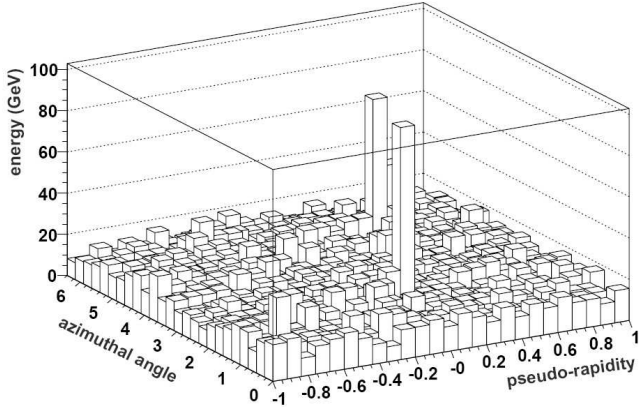}
\vskip -0.5 cm
\caption{Pseudorapidity-azimuthal angle plot of Pb-Pb event at LHC energy 
with two 100 GeV jets generated with HIJING and PYTHIA event generators. From
\cite{lhcp2}.}
\vskip -0.5 cm
\label{fig:lhc2}
\end{center}
\end{figure}

One of the first and easiest measurements at ALICE would be that of
the charged-particle multiplicity in the mid-rapidity region. Particle
production models and simple fits which are in agreement with the AGS,
SPS, and RHIC data on this quantity differ substantially from each
other when extrapolated to the LHC energy, as shown in
Fig. \ref{fig:lhc1}.  Thus this simple ``first-day'' measurement will
test our understanding of the physics of multiparticle production. The
charged-particle multiplicity provides a handle on the initial entropy
production; the latter quantity is a necessary input in the
hydrodynamic evolution of the produced matter.

Another relatively simple measurement at ALICE would be that of the
elliptic flow $v_2$ which has played a crucial role at RHIC (sec. {\bf
3.2}). The initial energy density (eq. (\ref{bjform})) as well as the
QGP lifetime are predicted to be higher at LHC than those at
RHIC. This is expected to raise the value of $v_2(p_T)$. On the other
hand, the increased radial flow at LHC is expected to lower
it. (Recall the discussion on mass ordering in sec. {\bf 3.2}.) The
net effect on $v_2(p_T)$ depends on the mass of the hadron:
Minimum-bias $v_2(p_T)$ for pions (protons) is expected to be higher
(lower) at LHC than at RHIC, at low $p_T$; see Eskola et al. in
\cite{armesto}. Prediction by Kestin and Heinz is that $v_2(p_T)$ at a
fixed impact parameter will be smaller at LHC than at RHIC, for pions
as well as protons \cite{armesto}. However, $p_T$-integrated elliptic
flow is expected to be higher for all hadrons due to the increased
relative weight at large values of $p_T$.

In sec. {\bf 3.5} we have quoted the values of $T_{\rm ch}$ and
$\mu_B$ for the SPS and RHIC energies. The latest predictions for LHC
are $T_{\rm ch}=161 \pm 4$ MeV and $\mu_B=0.8^{+1.2}_{-0.6}$ MeV
\cite{armesto}.

Hard processes: Cross sections for the production of heavy flavours,
$\sigma_{c{\bar c}}$ and $\sigma_{b{\bar b}}$, are expected to be
about 10 and 100 times larger at LHC than at RHIC. Cross sections for
the production of jets with transverse energy in excess of 100 GeV are
expected to be several orders of magnitude higher. Jet-photon events
will also be abundant.  Figure \ref{fig:lhc2} displays the capability
of ALICE to reconstruct the high-energy jets at LHC in spite of the
large soft-hadron background.  Thus it would be possible to make
detailed differential studies of heavy-quarkonium production,
open-charm and open-beauty production, jet quenching, etc. at LHC
\cite{armesto}. It will also be possible to study quark mass
dependence and colour charge dependence of the energy loss of a parton
as it traverses the medium.

Thus LHC promises to be a valuable tool to test our models of
ultrarelativistic heavy-ion collisions and deepen our understanding of
QCD. For details, see \cite{alice}.



\end{document}